# Investigation of drop coalescence via transient shape evolution: A sequential event based approach

Krishnayan Haldar[a], Samarshi Chakraborty[a], Manas Ranjan Behera[a], Sudipto Chakraborty[a]*

[a] Department of Chemical Engineering, IIT Kharagpur, Kharagpur-721302, West Bengal, India

*Corresponding author:
Sudipto Chakraborty,
Department of Chemical Engineering,
IIT Kharagpur -721302
India.
Email: sc@che.iitkgp.ernet.in
Phone (office): +91 - 3222 – 283942

# ABSTRACT

While a drop of liquid is placed on another liquid surface, two possible coalescence outcomes are observed. The parent drop bounces several times, floats and then disappears within the liquid pool without producing daughter droplets. This is called complete coalescence. Another outcome is the generation of secondary droplets from the primary drop itself. This is called partial coalescence. Repetitions of such phenomenon as a successive self-similar event is also known as coalescence cascade. In a nutshell, complete coalescence is governed strongly by swallowing mechanism whereas partial coalescence is attributed to slippage mechanism and solutal Marangoni flow. Here we use high speed camera and witness that water drop coalesces completely after impacting on pool of liquid whereas drop of non-ionic surfactant (TWEEN 20) coalesces partially. We also observe that number of daughter droplet generation is a strong function of surfactant concentration. Here we utilise the images to elaborately explain both the phenomenon in a sequential event based approach using the basic understanding of physics. We observe the transient evolution of drop morphology during each stage and find shape wise resemblances with many common objects. Our study reveals some interesting insight on the intermediate mechanisms and enlightens distinctive features of both the phenomena. Here we propose for the first time that additional instability due to slippage of rising water layer from the surfactant drop and upward thrust due to Laplace pressure helps in formation of secondary droplet.

# I. INTRODUCTION

Many natural occurrence like falling of rain drop [1], formation of cloud mist [2], vortices [3], mixing in microfluidic devices [4], foam formation [5] etc. are assumed as examples of coalescence process many years ago. Since, many researchers have looked into it theoretically and experimentally to have an insight of it. There are sequences of stages which occur before the drop merged with the bulk. The invention of high speed camera has enabled researchers to observe the complex mechanism of drop coalescence. The bigger drop falls with a low impact velocity [6], forms a crater on the water surface and then sinks within it quickly. Accordingly, due to opposite reaction force of the liquid pool the drop rebounds from the crater. Then the drop repetitively bounces until it gets stabilized on water surface. It was found from the study of Charles & Mason [7] that the drop rests over a very thin layer of air entrapped between drop and bulk interface. This entrapped air has a commanding role on duration of drop floatation on water surface [8]. Subsequently, this weak layer of air drained away to aid in merging of drop and bulk liquid [9]. Now, the liquid mass of drop has the chance to flow into the bulk water because of air drainage. As a result a capillary wave is generated across the periphery of the drop [10]. It was also suggested that due to sudden water evacuation the droplet volume gets reduced, therefore creating a narrow neck immediately. The completion of narrowing increased the chance of pinch off. This process of generation of secondary droplets in six to eight steps due to partial coalescence is called coalescence cascade [11].

The initial drop shape and the consecutive shape change from prolate to oblate or vice versa of falling drop played a role in generation of secondary droplets during partial coalescence [12]. It was also suggested by the researchers, that dual secondary drops are generated from prolate shaped drops while only single drop is generated for oblate drops. Blanchette & Bigioni [13] showed that a race between rates of vertical and horizontal collapse of drop played a role in partial coalescence. If the horizontal collapse rate succeeds over the vertical collapse rate, then it can produce a secondary droplet.

Some researchers[11, 14-16] have tried to explain dynamics of partial coalescence in absence of any surface active ingredients. Surface properties like elasticity, longitudinal viscosity and surface tension of surfactant influence in squeezing of drop at interface [17]. Chen et al. [15] found that partial coalescence is mostly dependent on inertia and interfacial tension. They demarcated the whole phenomena into three regimes considering the effect of gravity and viscosity. It was observed from the study of Thoroddsen & Takehara [11] that the phenomenon is controlled by surface tension and inertia of drop while the viscosity has no contribution. The higher surface tension value of the reservoir fluid than the drop is a favourable condition for partial coalescence [18]. It was observed from numerical results of Blanchette & Bigioni [19] that low viscous drops will be able to generate secondary droplets. Even the vertical velocity of secondary droplet after getting pinched off is also dependent on viscosity. Numerical results of Ray et al. [20] proved that partial coalescence is governed by inertia and interfacial tension forces while it is suppressed by viscous and gravity forces. Also the neck oscillation has a significant contribution in conversion from complete to partial coalescence. Martin et al. [21] have described recently the influence of surfactant concentration on forming coalescence cascade. The viscosity and the elasticity of surface tension of the drop play a major role behind this. Gilet et al [22] studied coalescence process by impinging drop at the interface of two different liquids. They suggested the process of coalescence has been influenced by viscosities and density differences of both the fluids.

Hence lot of attempts have been made in recent past to explain coalescence phenomena; still many aspects of it remains opaque to scientific community. Here we found some new stages happening intermediately and going to propose a clear cut theory on it. We performed experiments using drop of water and non ionic surfactant impinged on bulk of water. We observed complete coalescence for water drop and partial coalescence for surfactant drop. We also found a very interesting phenomenon that the number of droplet generation is dependent on surface tension of the impinging drops.

## II. EXPERIMENTAL SETUP

A schematic diagram of the experimental set up has been shown in Fig.1 where a transparent glass beaker (Borosil) of 11.5 cm height and outside diameter of 86.8 mm and inside diameter of 78.74 mm is used for all these experiments. Distilled water has been filled 3/5$^{th}$ of total beaker volume to create the liquid pool. The beaker is placed on a laboratory jack which height can be adjusted by rotating a screw-handle. A micro syringe (Rame−Hart 100-10-12-22) having needle dimensions of 0.711mm outer diameter and 0.406 inner diameters has been clamped with a stand above the liquid pool. The distance between needle tip and liquid surface has been kept 5 mm by adjusting the laboratory jack. The diameter of the drop depends on the needle diameter and liquid properties. This diameter is calculated by dropping 1 ml of volume for each solution and counting the number of drops. The height difference of 5 mm ensures gentle placing of liquid drop on air−liquid interface and prevention of splashing. A high−speed digital camera (Phantom V 7.3, Vision Research Inc. USA) is used to capture the drop impact dynamics. The video has been recorded in 512 x 384 resolutions at 14035 frames per second with exposure time of 69 μs. The image is maximum macro focused in accordance with 85X zoom and aperture opening between 5.6-8 mm. For all the experiments scale has been fixed at 0.056 mm/pixel. The camera is connected with a computer having in built software (Phantom PCC) by an Ethernet cable.

The PCC software is kept opened during experimentation to observe the live video. The camera is kept in slight inclination of 5$^o$ to capture the interfacial behaviour. A LED light source (24 W) is kept behind the beaker (backlighting) with a glass diffuser in front of it. The diffuser is used to prevent over lighting which can make image unclear. The captured video is recorded in a data recording computer. The camera, computer and light source is connected to a common switch board. Trigger button of the PCC software is pressed as soon as the drop detaches from the needle. There is an extra pre-triggering option in the software which compensates minor time delay between event occurrence and switching of trigger. The recorded video is saved in .avi format and then cropped to eliminate unwanted parts. Then it is converted to images using free software Free Video to JPG Converter (DVDVideoSoft Ltd.). The images are analysed frame by frame to track the intermediate events of the phenomenon and processed accordingly. Necessary actions have been taken to avoid all possible experimental errors like needle and beaker vibration etc.

We have used pure water and four different concentrations of non-ionic surfactant solution (TWEEN 20). TWEEN 20 (Polyoxyethylene sorbitan monolaurate; CAS Number: 9005-64-5; density 1.100−1.110 gm/cm$^3$) has been procured from Merck, India. Distilled water of pH 5.6 has been used to prepare all solutions. The physical properties of each each solution have been displayed in Table 1. Surface tension, viscosity, specific gravity of solutions is measured using Du Nouy ring tensiometer (Testing Instruments, India), Parallel plate type viscometer (Brookfield DV-II+Pro) and Relative Density Bottles of 25ml, I/C Teflon Stopper (1625) (Borosil Glass Works Ltd.)

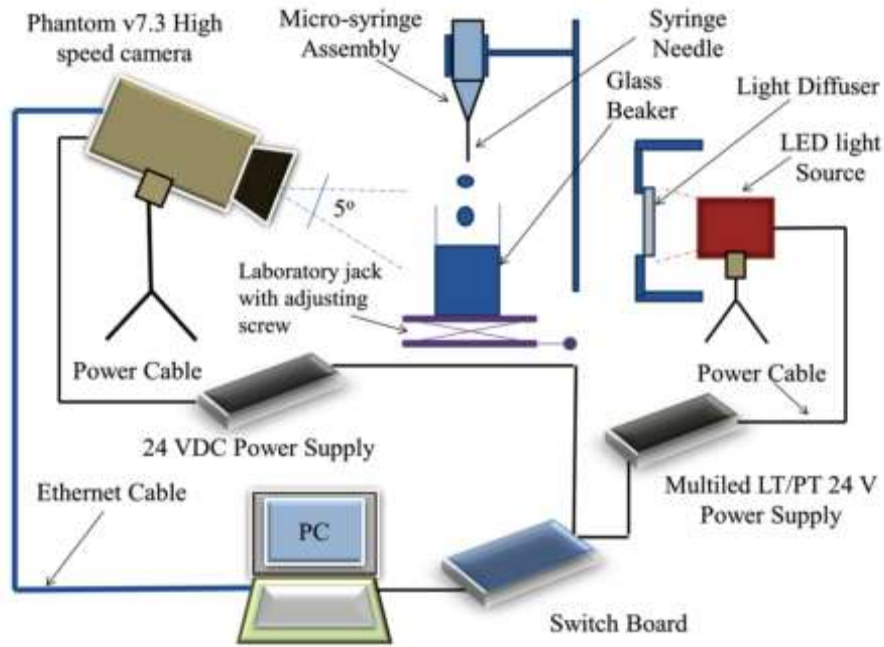

**FIG. 1. Schematic diagram of experimental set up**

**Table I: Diameter and physical properties of fluids of different concentrations used for these experiments.**

| Concentration (ppm) | Surface tension (dyne/cm) | Drop diameter (mm) | Viscosity (mPa-s) | Specific Gravity |
|---|---|---|---|---|
| 0 | 71.4 | 2.98 | 1.00 | 0.99340 |
| 36.5 | 44.1 | 2.83 | 0.90 | 0.99376 |
| 73 | 39.6 | 2.77 | 0.85 | 0.99444 |
| 730 | 38.4 | 2.59 | 0.88 | 0.99492 |
| 3650 | 37.7 | 2.54 | 1.10 | 0.99560 |

## III. RESULTS AND DISCUSSION

### A. Complete Coalescence

While a drop of water is impinged from a needle, it coalesces completely in the bulk liquid (water). Here, we explain the coalescence behaviour of the parent drop by dividing in nine physical stages. Images are obtained from captured high speed video and analysed frame by frame.

1. **Floating of droplets**: The liquid drop of water is gently impinged on the bulk of water and it is observed that the drop rests over the surface before sinking into the bulk (Fig. 2). This time duration is known as floatation time/ residence time. It is known from the literature that a thin layer of entrapped air provides the necessary cushion to make the drop float [8]. The drop floats for 119 ms in this case.

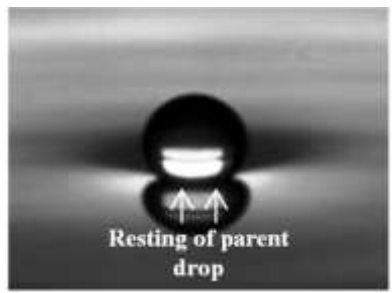

**FIG.2. Floatation of impinging water drop**

2. **Rupture of Air film**: The layer of air is very thin so it is quite unstable in nature and has possibility of getting ruptured. The rupture is due to the attractive Van der Waals interaction between the two interfaces of drop-air and liquid-air [23, 24] when its thickness is reduced below 100 nm. At first a hole is generated at the point of rupture (Fig. 3-a) when two interfaces merge together. Then the hole grows in size (Fig. 3-b) and propagates horizontally along the contact line (Fig. 3-c). As soon as the hole starts to propagate the drop bottom also merges with the bulk surface. Finally it is dissipated completely (Fig. 3-d) to form a whole entity of drop and bulk. This event continues for 0.355 ms.

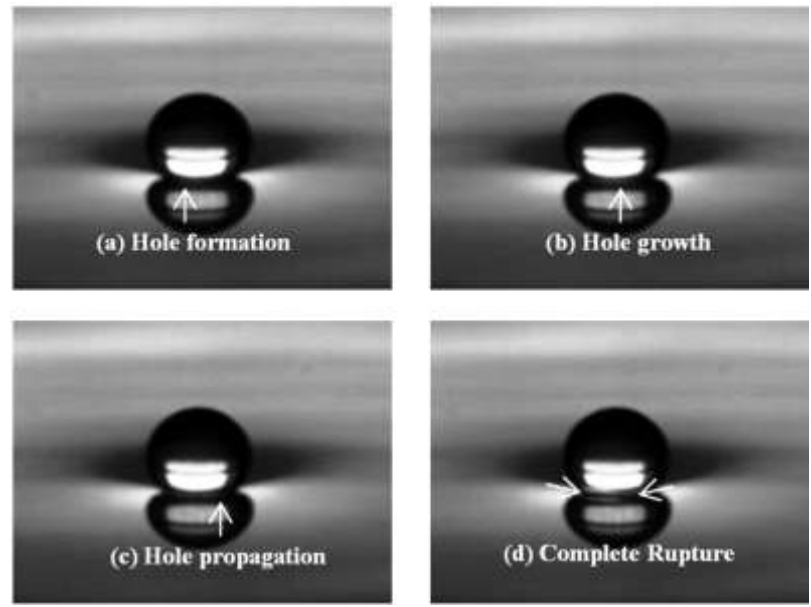

**FIG.3. Rupture of air film entrapped between drop and liquid surface. Snapshots obtained from high speed movies showing the process for t = 0 ms (a), 0.071 ms (b), 0.142 ms (c), 0.355 ms (d).**

3. **Rise of capillary wave**: The rupture of air film generates a surface capillary wave which is able to rise up along the drop surface. The direction of movement of this wave is radially outwards along the initial flat interface [20]. The following figures (Fig. 4-a–c) represent the dynamics of capillary wave rise. The changing location of white area signifies the movement of wave front. This event continues for 0.639 ms. which consists of corresponding sub events like generation, rise and final rise of the capillary wave.

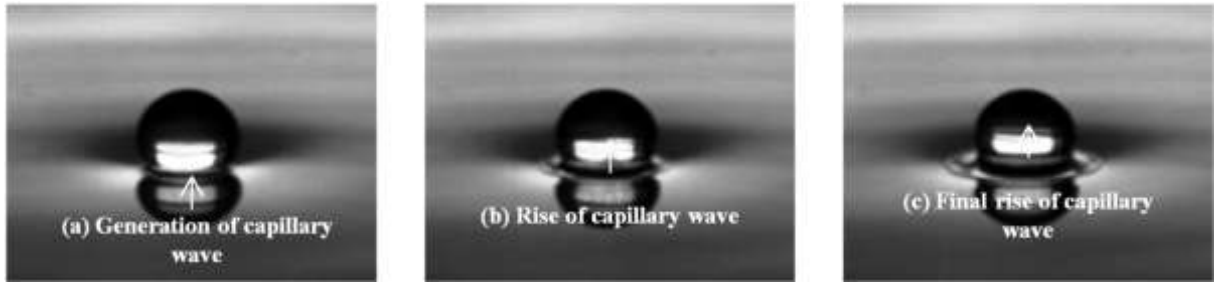

**FIG.4. Dynamics of capillary wave rise. Snapshots obtained from high speed movies showing the process for t = 0 ms (a), 0.355 ms (b), 0.71 ms (c).**

4. **Creation of engulfing water layer**: The following stage is supposed to be creation of layer of liquid from the bulk which will have tendency to engulf the drop. This stage is proposed by us for the first time which is the direct consequence of capillary wave rise. The energy carried by capillary wave is responsible for generation (Fig. 5-a) and rise of engulfing water layer. This layer will crawl over the drop gradually (Fig. 5-b) and will reach finally at the drop apex (Fig. 5-c). The interim shape change of drop takes place at 0.639 ms while final rise of engulfing water layer takes place at 1.207 ms. The final image shows a distortion of the original spherical form and a slight prick at the drop summit.

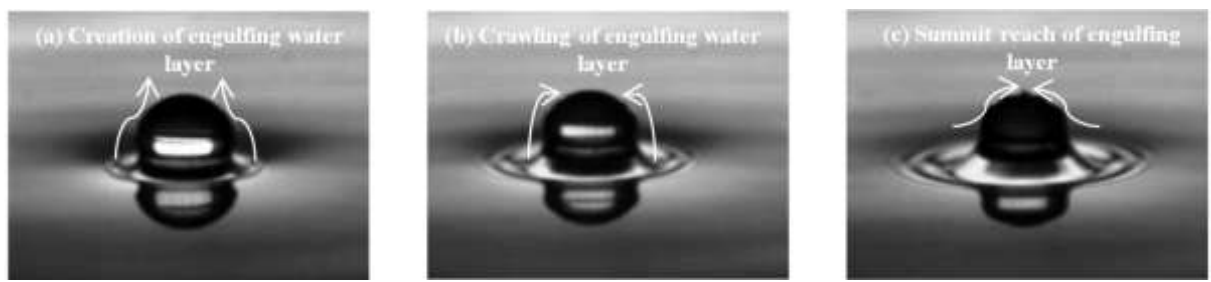

**FIG.5. Behaviour of engulfing water layer. Snapshots obtained from high speed movies showing the process for t = 0 ms (a), 0.639 ms (b), 1.207 ms (c).**

5. **Drainage of drop**: The sequential rise of capillary wave and engulfing water layer both impart a downward pressure on the drop (Fig. 6-a). This results in gravity driven flow from the drop through the interface. The drop takes different shapes and consequent physical sub−events happened intermediately. All the sub−events are discriminated by observing the shape change of drop apex. The result of applied pressure is dip down of peak (Fig. 6-b), and peak flattening (Fig. 6-c). Suddenly another secondary peak starts to rise from that site (Fig. 6-d) and quickly gets flattened out (Fig. 6-e). The gravity driven flow is completed at 0.781 ms when we observe the total suppression of secondary peak (Fig. 6-f). It is prominent that not the whole drop rather a part of it is drained out by the action of gravity.

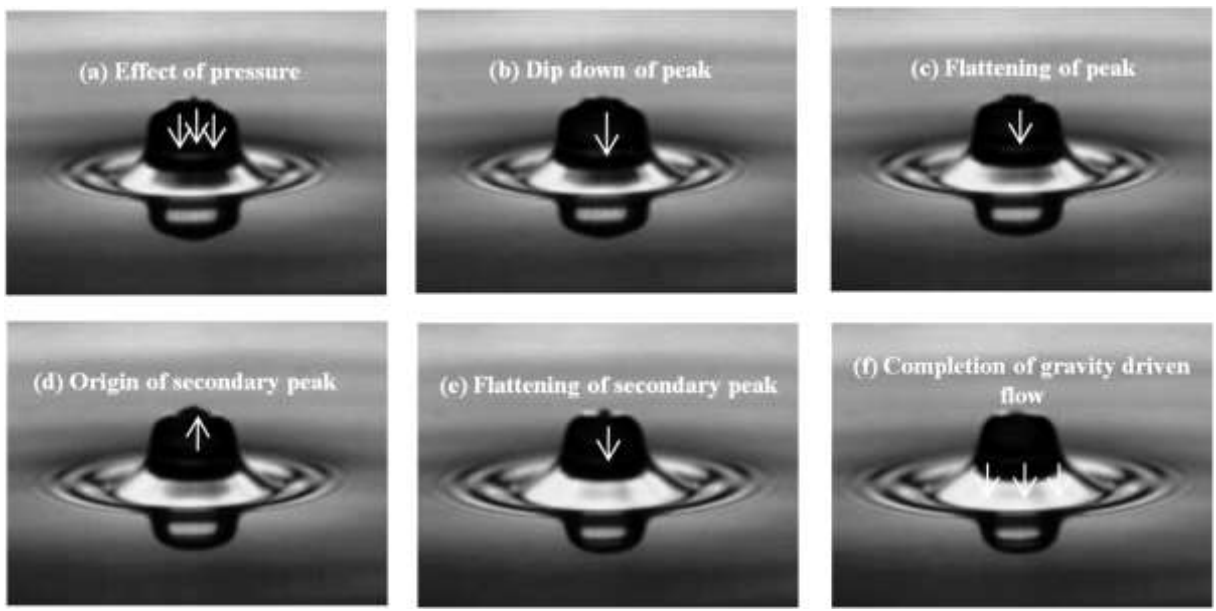

**FIG.6. Flow of liquid through ruptured interface. Snapshots obtained from high speed movies showing the process for t = 0 ms (a), 0.071 ms (b), 0.284 ms (c), 0.355 ms (d), 0.71 ms (e), 0.781 ms (f).**

6. **Swallowing**: The major reason of complete coalescence is the swallowing stage which takes place just after the drainage of the drop. The continuous mass loss because of downward flow makes it easier to get it swallowed by engulfing layer. The engulfing water layer will rise above the drop and will try to swallow it completely. This event continues for 0.426 ms where engulfing layer of water gulps the whole drop of water. The visible change in drop shape is observed and demonstrated in the following figures ((Fig. 7 a-c). The top notch part of the drop changes its shape. Initially it was a bit flat whereas at later stages it converges to a pin like profile.

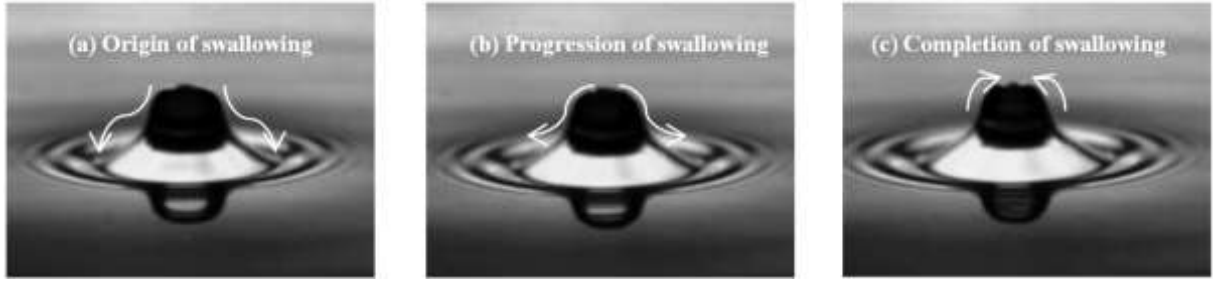

**FIG.7. Dynamics of drop swallowing by engulfing water layer. Snapshots obtained from high speed movies showing the process for t = 0 ms (a), 0.213 ms (b), 0.426 ms (c).**

7. **Narrowing of swallowed drop**: Another significant shape change of drop occurs at this event. The drop acquires various shapes as displayed below. The drop is narrowed from both sides as a consequence of Laplace pressure ( $\Delta P = P_{in} - P_{out} = 2\gamma / R$ ) generated upward thrust. Laplace pressure is the difference between inside and outside pressure of the drop respectively across curved liquid−air interface which originates because of surface tension. The outward thrust will force the drop gets narrowed at interface (Fig. 8-a). As it is narrowed from both sides, a neck is erupted vertically (Fig. 8-b) from the drop (0.284 ms). Slowly its neck is elongated (Fig. 8-c) and a nipple like shape is appeared (Fig. 8-d). This nipple seems to get pinned at its interface with bulk (Fig. 8-e). Pinning is a physical process where the body of the drop gets attached with the liquid pool. This stage lasts for 1.42 ms when the swallowed drop narrowed totally (Fig. 8-f).

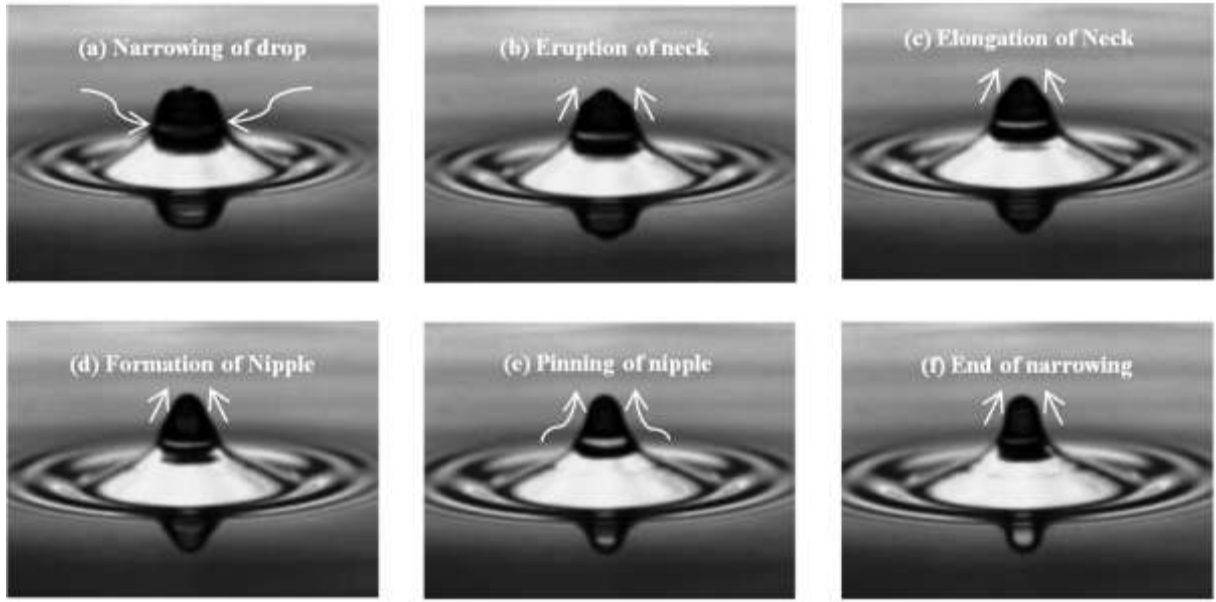

**FIG.8. Morphology of drop during narrowing. Snapshots obtained from high speed movies showing the process for t = 0 ms (a), 0.284 ms (b), 0.568 ms (c), 0.852 ms (d), 1.136 ms (e), 1.42 ms (f).**

8. **Gradual collapse of liquid drop**: A competition will act between outward Laplace pressure and downward engulfing pressure. The extent of engulfing pressure suppresses the effect of thrust of Laplace pressure completely. Therefore, the drop gets reduced in height and starts to collapse within the pool. The mechanism is schematically illustrated in Fig. 9.

   It takes a form of small pillar at the end of squeezing (Fig. 10-a). The pinning effect is solely responsible for creating a pillar like formation. The height of the drop pillar slowly reduces (Fig. 10-b), pins (Fig. 10-c) and then starts to collapse (Fig. 10-d). This collapse is due to cumulative effects of engulfing, swallowing and downward gravity driven flow. Actually the drop sinks within the basin (Fig. 10-f) and the whole event takes much higher time (3.124 ms) for completion.

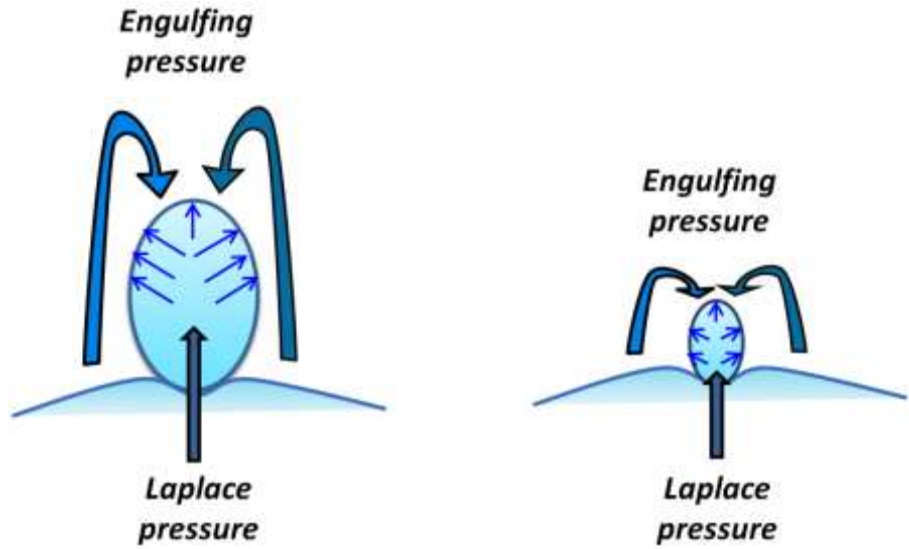

**FIG.9. Schematic illustration of competitive effect between engulfing pressure and upward thrust of Laplace Pressure leading to drop collapse**

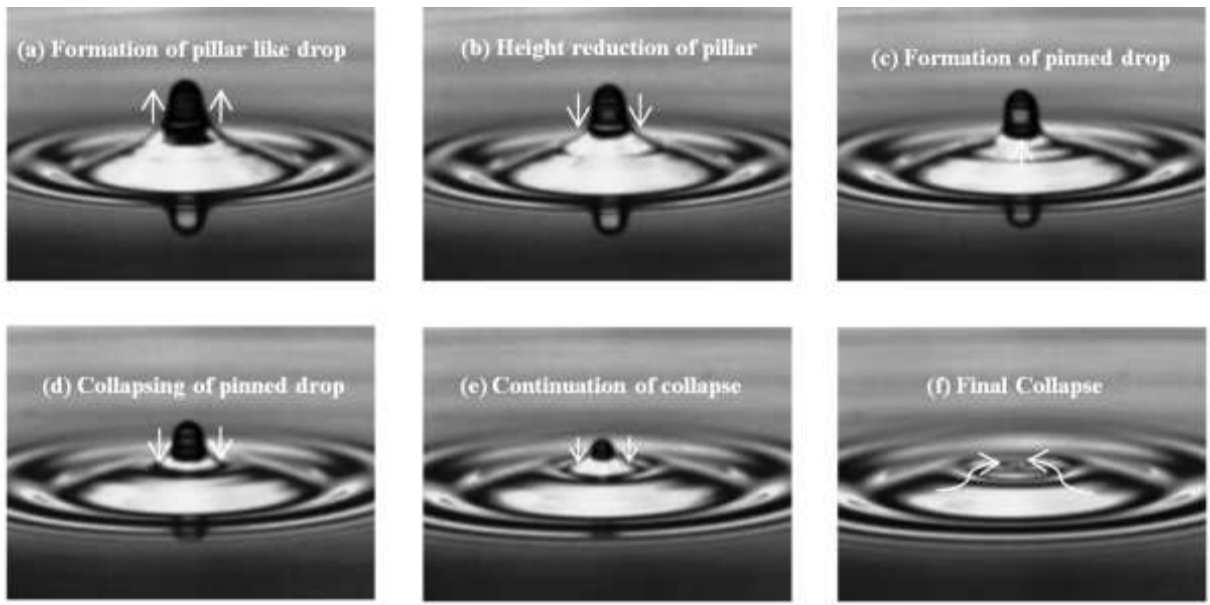

**FIG.10. Collapse of liquid drop after narrowing. Snapshots obtained from high speed movies showing the process for t = 0 ms (a), 0.355 ms (b), 1.065 ms (c), 1.775 ms (d), 2.485 ms (e), 3.124 ms (f).**

9. **Collapse of drop**: This image is taken immediately after the final collapsing stage. The just risen peak of the column instantaneously falls to produce complete coalescence (Fig. 11). Hence, no daughter droplet generation is possible for this situation.

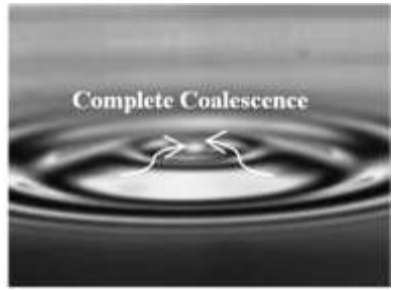

**FIG.11. Collapse of water drop within pool of water.**

B. **Partial Coalescence**

While a drop of surfactant solution is impinged from a needle, it coalesces partially for four to five times (depending on concentration) on the bulk liquid. Here, we explain the first coalescence behaviour of the parent drop by dividing in ten physical stages and assume rest of the secondary drops follows similar coalescence behaviour. Images are obtained from captured high speed video and analysed frame by frame.

1. **Resting of surfactant covered drops**: The surfactant laden drop is gently deposited on the liquid pool. It tends to float on the surface for quite some time before the start of coalescence (Fig. 12); the nature of floatation on the entrapped air layer is similar to the phenomenon explained for resting of water drop. Resting of drops is the first and foremost stage of coalescence. The drop floats for 123.05 ms for this case.

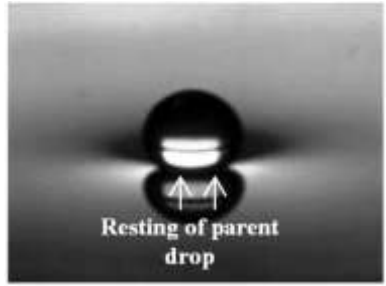

**FIG.12. Floatation of impinging surfactant laden water drop**

2. **Rupture of entrapped air layer**: The rupture of entrapped thin layer of air ruptures in the similar fashion as described earlier. Initially a hole is generated (Fig.13-a) which grows (Fig.13-b), propagates horizontally (Fig.13-c) and then finally completely dissipates away (Fig. 13-d). Thus the drop merges with the liquid pool to make a complete entity. The total time taken for rupture of air film for this case is 0.355 ms.

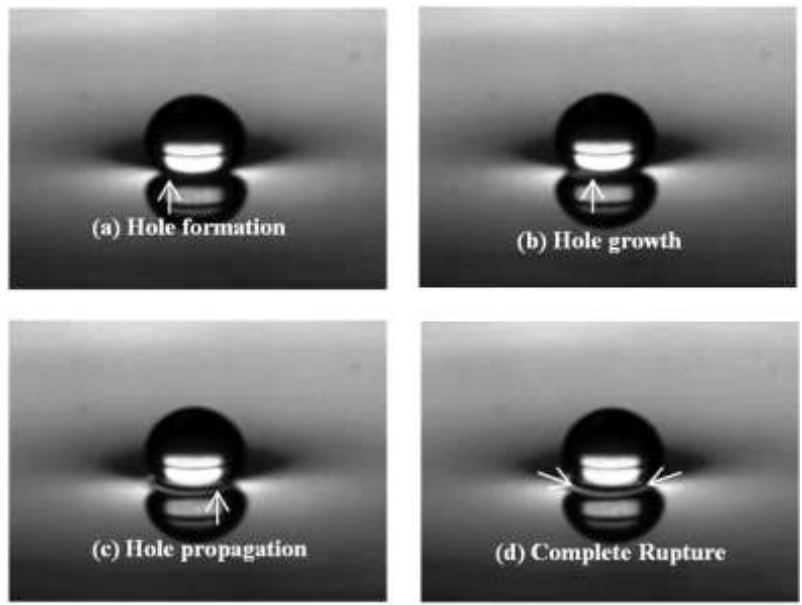

**FIG.13. Rupture of air film entrapped between drop and liquid surface. Snapshots obtained from high speed movies showing the process for t = 0 ms (a), 0.071 ms (b), 0.284 ms (c), 0.355 ms (d).**

3. **Rise of capillary wave**: The subsequent effect of capillary wave rise due to rupture of film also follows the same mechanism like it happens over water drop. A vertical capillary wave generates (Fig. 14-a), rises, crawls (Fig. 14-b) over the drop body and finally it ends (Fig. 14-c). The wave propagation is observed in the following images. The important thing is that the drop shape doesn't change at all while a reflecting layer just changes its location across the drop periphery. The capillary wave takes 0.568 ms to rise across the drop.

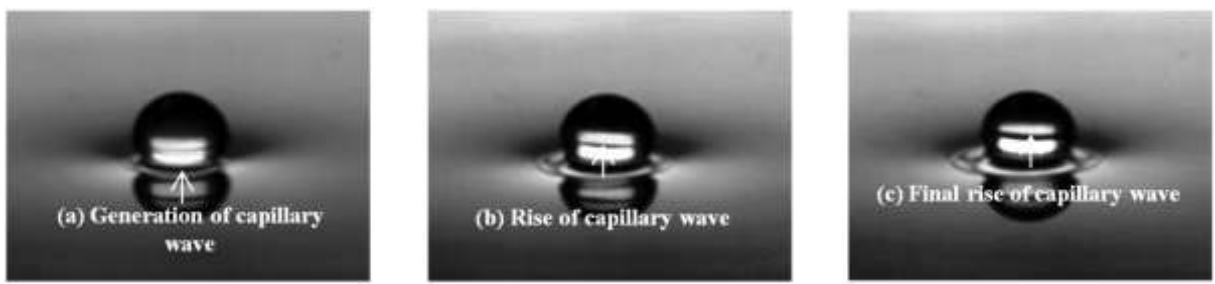

**FIG.14. Process of formation and rise of capillary wave. Snapshots obtained from high speed movies showing the process for t = 0 ms (a), 0.355 ms (b), 0.568 ms (c).**

4. **Rise of engulfing water layer**: Similarly, as explained for water drop incident a thin layer of water from the bulk is created which has a general affinity to swallow the drop. The rise of water layer over the drop is clearly visible from the following images. The drop shape is changed notably at 1.207 ms due to summit reach of engulfing layer (Fig. 15 a-c).

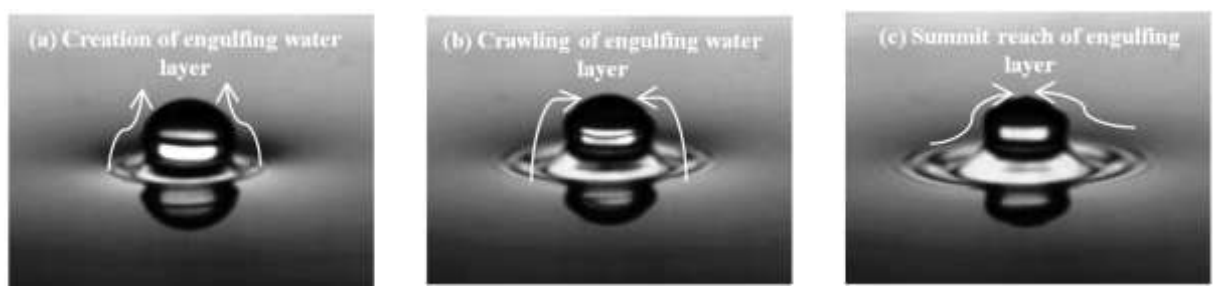

**FIG.15. Behaviour of engulfing water layer. Snapshots obtained from high speed movies showing the process for t = 0 ms (a), 0.71 ms (b), 1.207 ms (c).**

5. **Gravity driven flow along with diffusion:** The rise of engulfing water layer will pressurize the drop fluid to drain down (Fig. 16-a). Although the drop will drain due to action of gravity, there will also be an action of concentration driven Marangoni flow or solutal Marangoni effect[25, 26]. It is quite evident that concentration varies at drop-bulk interface because of presence of surfactants in the drop. Hence, a surface tension gradient acts at interface which aids in origin of Marangoni flow. This flow involves mass transfer in the form of diffusion through the interface. This is the stage which differentiates from the water drop draining. Similar observations of primary peak dipping (Fig. 16-b), primary peak flattening (Fig. 16-c), secondary peak generation (Fig. 16-d) and its flattening (Fig. 16-e) are observed like in the previous case. The total time taken to complete (Fig. 16-f) this gravity driven and concentration driven flow is 0.852 ms.

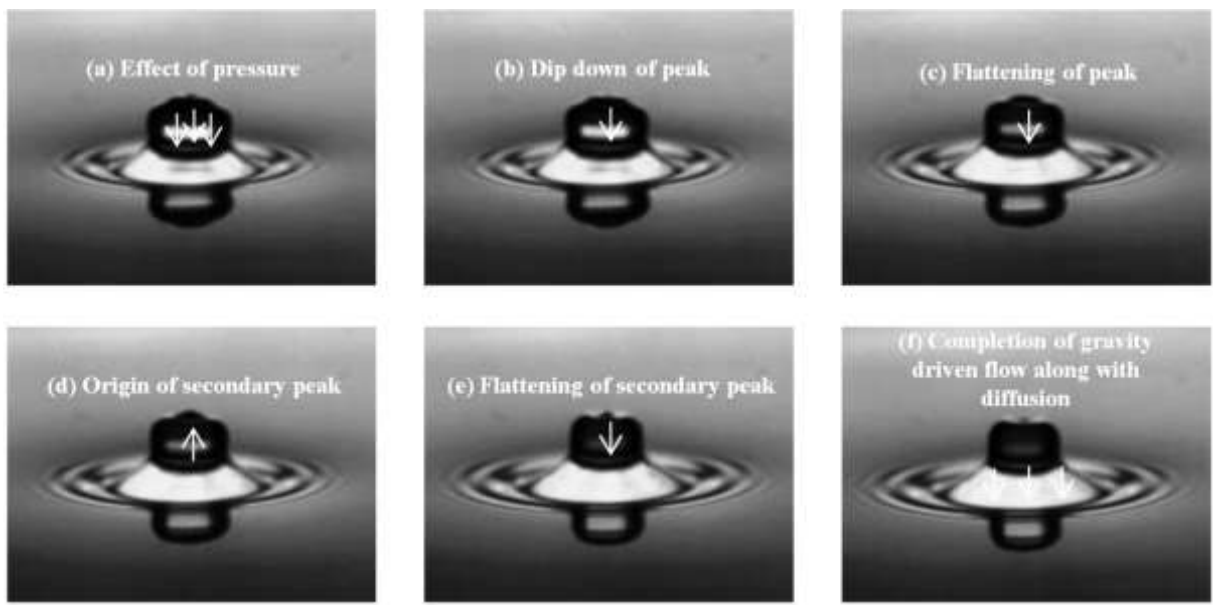

**FIG.16. Flow of liquid through ruptured interface. Snapshots obtained from high speed movies showing the process for t = 0 ms (a), 0.071 ms (b), 0.355 ms (c), 0.426 ms (d), 0.781 ms (e), 0.852 ms (f).**

6. **Slippage of water layer:** Although the engulfing water layer will rise over the drop, still it will not gulp the underlying drop like in case for water drop. Rather it will slip from the surface of the drop across the periphery. Here we report the occurrence of slippage mechanism for the first time in context of partial coalescence. The mechanism of slippage from the surface is explained schematically in the Fig. 17. The surfactant molecules orient itself in the drop air interface by keeping the hydrophobic tail hanging in the air while the hydrophilic head inserted within the bulk. The moment the layer of engulfing water climbs around the surface, it will be repelled by wagging hydrophobic tail. Therefore, the layer can't climb completely and slide from the surrounding of the drop. The slippage of engulfing water layer is the determining factor for partial coalescence. The images obtained from high speed camera are displayed (Fig. 18 a-c) to calculate the time of the event (0.568 ms). The periodic suppression and opening of peaks imply the progress of slippage.

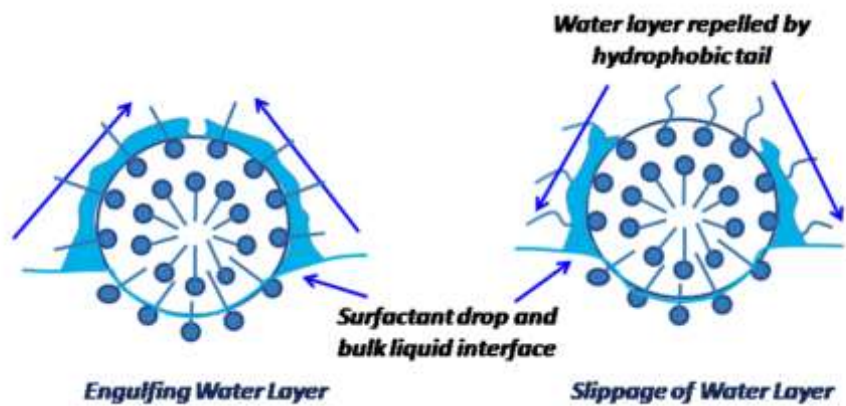

**FIG.17. Schematic illustration of hydrophobic repulsion by tail to initiate slippage**

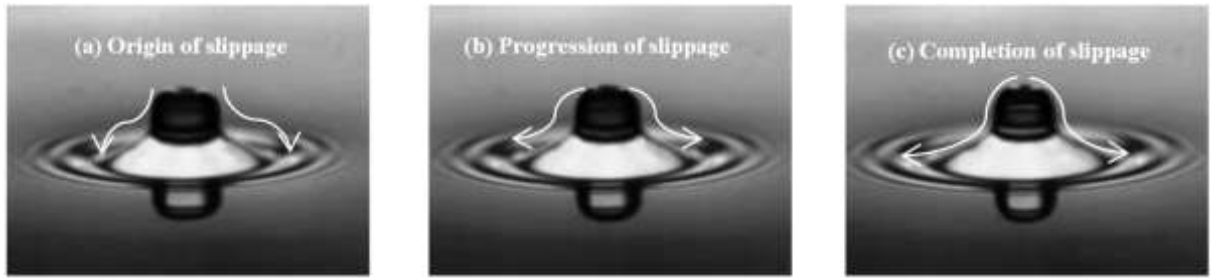

**FIG.18. Dynamics of slippage mechanism of engulfing water layer over the drop. Snapshots obtained from high speed movies showing the process for t = 0 ms (a), 0.284 ms (b), 0.568 ms (c).**

7. **Formation of column**: The stage which follows the slippage is also contributing to the partial coalescence. In this stage drop takes a shape of column. A tip is erupted from the drop vertically (Fig. 19-a), which elongates further (Fig. 19-b) and finally the drop appears as a column (Fig. 19-c). The whole event takes place for 1.136 ms.

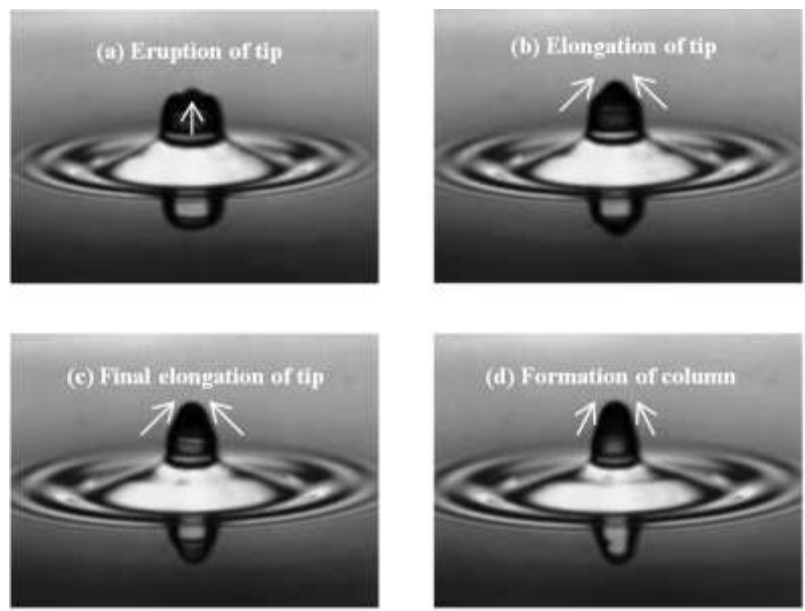

**FIG.19. Dynamics of column formation. Snapshots obtained from high speed movies showing the process for t = 0 ms (a), 0.426 ms (b), 0.852 ms (c), 1.136ms.**

8. **Column thinning:** The drop will slowly be reduced in mass because of solutal Marangoni flow. Marangoni flow will create an interfacial instability [27]. Also Laplace pressure ($\Delta P = P_{in} - P_{out} = 2\gamma / R$) which is the pressure difference between inside and outside drop pressure respectively comes will apply a thrust outwardly. These two actions will help in thinning of column (Fig. 20 a-c).

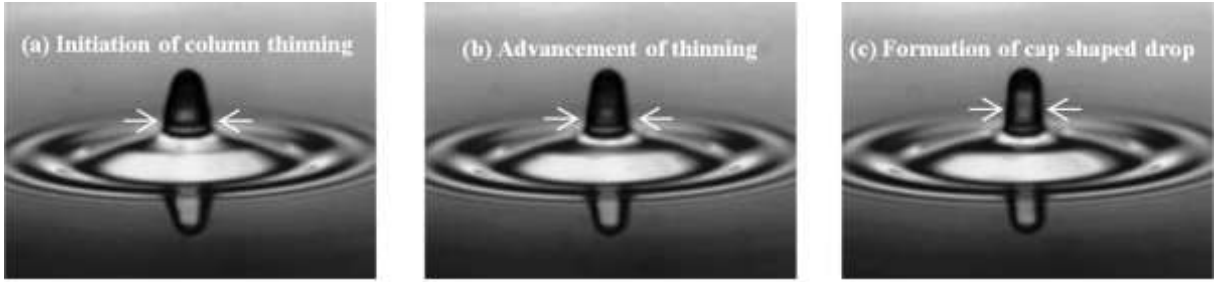

**FIG.20. Dynamics of column thinning. Snapshots obtained from high speed movies showing the process for t = 0 ms (a), 0.284 ms (b), 0.568 ms (c).**

9. **Neck thinning:** This stage is the longest among all and consists of various sub stages. As we explained earlier that Laplace pressure induced thrust will push the drop vertically upwards while the Marangoni stress induced flow will make it weaker at interface. These two acts as a tearing action for the drop which are schematically illustrated in Fig. 21.

The cap shaped drop will squash (Fig. 22-a) first to get narrowed down at neck (Fig. 22-b). After few milliseconds a concave interface (Fig. 22-c) will be formed at the junction of drop and bulk. Gradually the drop will take a shape of parachute (Fig. 22-d) and start to fall down (Fig. 22-e) along with the interface. The wider neck will be further crushed (Fig. 22-f) to form a thin strip. Then it will again change in the shape of balloon (Fig. 22-g). The balloon shaped drop will be deformed further (Fig. 22-h) to reach a stage of pinching off (Fig. 22-i).

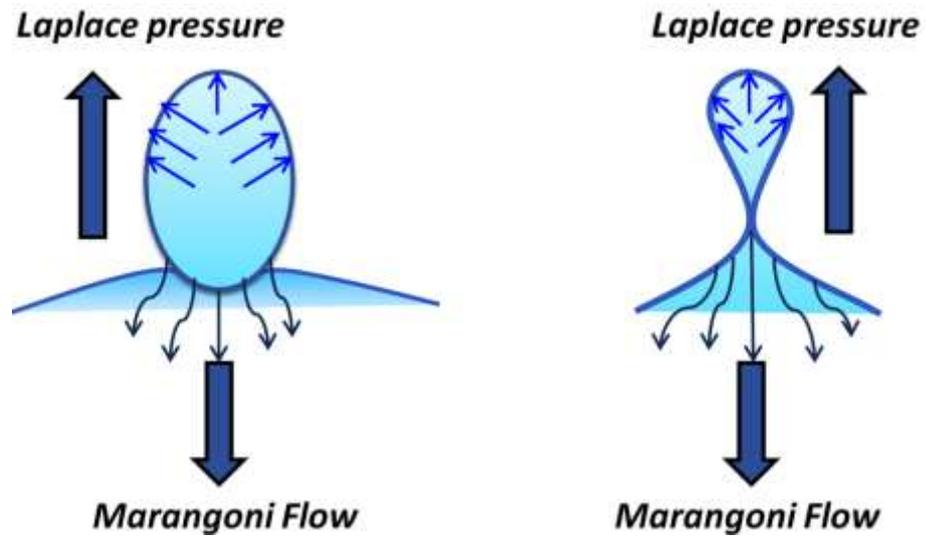

**FIG.21. Schematic illustration of effect of Laplace pressure and Marangoni flow generated instability on drop neck thinning.**

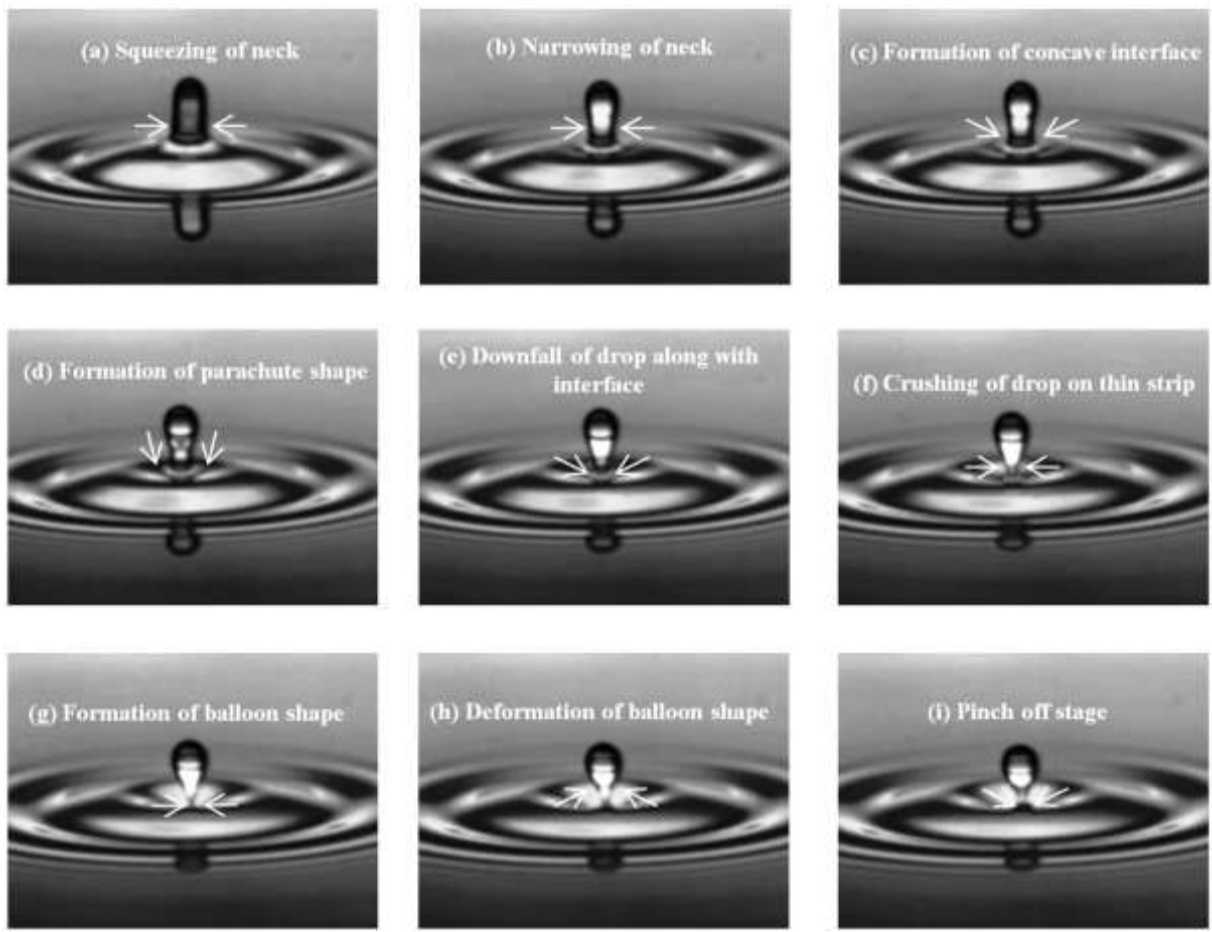

**FIG.22. Morphological evolution of drop during thinning. Snapshots obtained from high speed movies showing the process for t = 0 ms (a), 0.71 ms (b), 0.781 ms (c), 1.207 ms (d), 1.562 ms (e), 1.917 ms (f), 2.343 ms (g), 2.556 ms (h), 2.698 ms (i).**

10. **Detachment of daughter droplets**: The already pinched off drop detaches from the thin little strip and will give a secondary droplet (See Fig. 23). It pops from the surface and bounces similarly to give another set of partial coalescence.

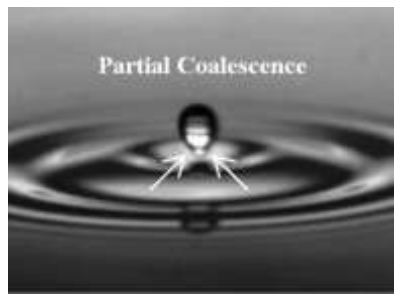

**FIG.23. Partial coalescence of surfactant laden water drop within pool of water.**

C. Coalescence Cascade

After impact, surfactant laden water drop coalesces partially i.e. jumps, floats and breaks up into another drop. This phenomenon of partial coalescence occurs in self similar manner repetitively in a series. The behaviour of coalescence cascade is observed for four times (36 ppm) and for five times (73, 730, 3650 ppm). Each of the outcomes of the cascade for surfactant concentration of 73 ppm is depicted in Fig. 24. (a−e). It is observed that drops are getting smaller in size at the end of each partial coalescence. The smallest drop (Fig. 24-e) coalesces completely in the same process as earlier explained. Similar images are obtained for drop of 36.5 ppm concentration only difference is its number.

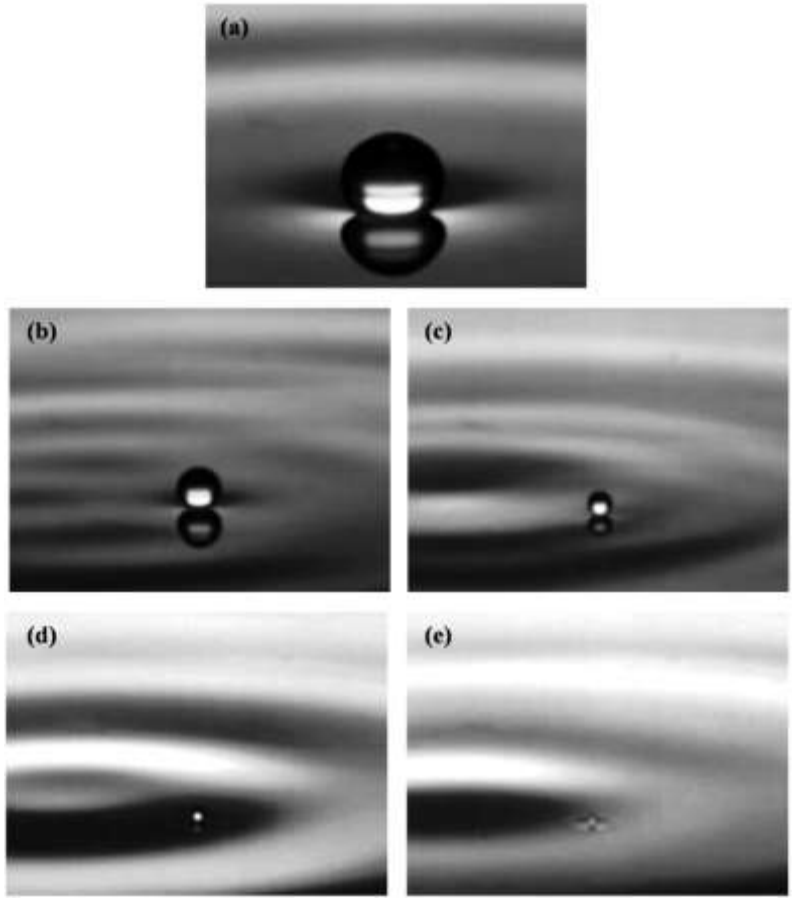

**FIG. 24. Successive generation of secondary droplets for surfactant drop (CMC) (a) Primary drop; (b) Secondary drop; (c) Tertiary drop; (d) Quaternary drop; (e) Pentanary drop.**

### D. Effect of concentration on coalescence time

An extensive frame wise analysis of the video helps in calculating time taken for each coalescence event. Figure 25 shows the variation of coalescence time for each of the drop at all surfactant concentrations. It is seen that coalescence time is getting reduced in succession along with reducing drop sizes. It is known from the expression of Laplace pressure, smaller drops (drops of reduced diameter) experience more thrust of Laplace pressure as it is inversely proportional with radius. Hence, secondary droplets require shorter time to coalesce.

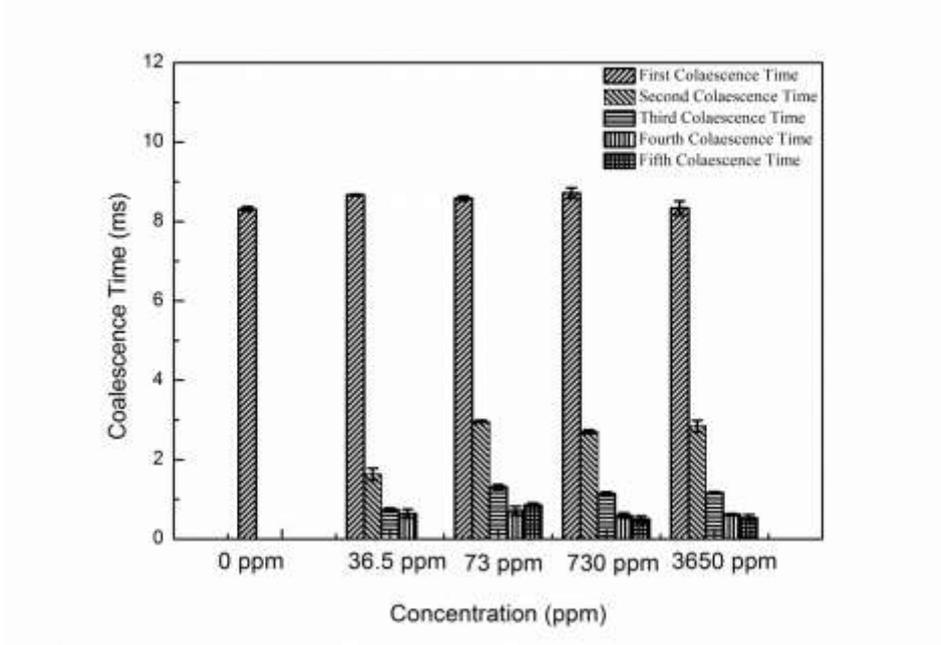

**FIG. 25. Effect of concentration on individual coalescence time**

## IV. CONCLUSION

We have performed experiments on liquid drop impacting on liquid pool by using water and four different concentrations of surfactant (TWEEN 20) solution. Using high speed photography we observe that upon impinging on a pool of water, a water drop doesn't break up into a secondary droplet whereas a surfactant laden drop can break into a series of daughter droplets. We split the whole phenomenon in distinct stages sustaining for definite time duration and narrate it qualitatively which are not found in existing literatures. Here we propose some new mechanisms influencing the coalescence phenomena by applying common knowledge of physics. Intermediately, we observe the change in drop shape in various sub-stages.

Two phenomena resulting from a liquid drop impinging on a liquid pool have been discussed elaborately by splitting in many stages occurring at definite time intervals. Complete coalescence of water drop takes place in nine stages whereas partial coalescence of surfactant laden drop takes place in ten stages.

We can summarise the whole events in few points:

(a) Initial four events (up to rise of engulfing water layer) are common for both the phenomena. Difference starts from fifth event for surfactant drop where a mass diffusion driven by concentration gradient along with downwards gravity driven flow occurs through interface.

(b) Complete coalescence is governed strongly by swallowing mechanism. Upward thrust due to Laplace pressure acts as the source of detachment but it is dominated completely by downward impact due to engulfing pressure. Marangoni stress is not acting in this case as solutal Marangoni flow is absent across the interface. Hence, engulfing pressure pushes the drop downwards. The drop remained pinned with the bulk to stay as a fluid body. This drop later gradually sinks within the pool to coalesce completely.

(c) Swallowing event is replaced by slippage of engulfing water layer for surfactant laden drop. We define slippage as repulsion of water layer by hydrophobic tail of surfactant molecule. Successively all the following stages are characteristically different contrary to complete coalescence. Slippage of water layer creates an additional interfacial instability outside along with solutal Marangoni flow which takes place through the interface inside the drop. Simultaneously, thrust generated because of Laplace pressure across the drop circumference pushes it upwards continuously. These both ways tearing action (Laplace pressure in upward direction and Marangoni flow in downward direction) leads to thinning and necking of drop. Hence, the drop will get detached easily from the interface.

(d) Physically surfactant covered drop is smaller in size than water drop owed to lower surface tension values [28] (See Table 1). Therefore, the parent surfactant drop always feels a stronger Laplace pressure driven upward thrust than water drop. Hence, presence of surfactant make drop more susceptible to partial coalescence.

(e) It is obvious that Laplace pressure increases with reducing drop size after each detachment. The detachment of drop at the end of partial coalescence makes further small droplet. Also certain amount of dilution increases its surface tension value. Hence, the upward thrust is also getting stronger which helps in further and faster detachment of droplets and occurrence in partial coalescence. We show that coalescence time is getting shorter for secondary droplets. Also, we suggest that drop can't be pinned at the interface with bulk because both the liquids are different. Hence, the flow due to concentration difference will resist the drop to get pinned rather make it prone to instability and detachment.

(f) We propose that the secondary daughter drop has reduced concentration gradient with the bulk liquid. However, it is sufficient for having Marangoni flow. Thus the partial coalescence will occur repetitively until concentration difference diminishes at the final smallest droplet, when it coalesces completely. At this instance drop concentration becomes similar to that of water. As we know surface tension becomes constant beyond CMC so the cascade of dilution will happen in self-similar manner for remaining surfactant concentration, at and beyond CMC. Similarly, the daughter droplet generation will be same in number which was shown earlier (See Fig. 25). This water like situation arrives at fifth stage of cascade for 73, 730, 3650 ppm but at fourth stage for 36.5 ppm surfactant concentration.

**LIST OF FIGURES**

FIG. 1. Schematic diagram of experimental set up

FIG.2. Floatation of impinging water drop

FIG.3. Rupture of air film entrapped between drop and liquid surface. Snapshots obtained from high speed movies showing the process for t = 0 ms (a), 0.071 ms (b), 0.142 ms (c), 0.355 ms (d).

FIG.4. Dynamics of capillary wave rise. Snapshots obtained from high speed movies showing the process for t = 0 ms (a), 0.355 ms (b), 0.71 ms (c).

FIG.5. Behaviour of engulfing water layer. Snapshots obtained from high speed movies showing the process for t = 0 ms (a), 0.639 ms (b), 1.207 ms (c).

FIG.6. Flow of liquid through ruptured interface. Snapshots obtained from high speed movies showing the process for t = 0 ms (a), 0.071 ms (b), 0.284 ms (c), 0.355 ms (d), 0.71 ms (e), 0.781 ms (f).

FIG.7. Dynamics of drop swallowing by engulfing water layer. Snapshots obtained from high speed movies showing the process for t = 0 ms (a), 0.213 ms (b), 0.426 ms (c).

FIG.8. Morphology of drop during narrowing. Snapshots obtained from high speed movies showing the process for t = 0 ms (a), 0.284 ms (b), 0.568 ms (c), 0.852 ms (d), 1.136 ms (e), 1.42 ms (f).

FIG.9. Schematic illustration of competitive effect between engulfing pressure and upward thrust of Laplace Pressure leading to drop collapse.

FIG.10. Collapse of liquid drop after narrowing. Snapshots obtained from high speed movies showing the process for t = 0 ms (a), 0.355 ms (b), 1.065 ms (c), 1.775 ms (d), 2.485 ms (e), 3.124 ms (f).

FIG.11. Collapse of water drop within pool of water.

FIG.12. Floatation of impinging surfactant laden water drop

FIG.13. Rupture of air film entrapped between drop and liquid surface. Snapshots obtained from high speed movies showing the process for t = 0 ms (a), 0.071 ms (b), 0.284 ms (c), 0.355 ms (d).

FIG.14. Process of formation and rise of capillary wave. Snapshots obtained from high speed movies showing the process for t = 0 ms (a), 0.355 ms (b), 0.568 ms (c).

FIG.15. Behaviour of engulfing water layer. Snapshots obtained from high speed movies showing the process for t = 0 ms (a), 0.71 ms (b), 1.207 ms (c).

FIG.16. Flow of liquid through ruptured interface. Snapshots obtained from high speed movies showing the process for t = 0 ms (a), 0.071 ms (b), 0.355 ms (c), 0.426 ms (d), 0.781 ms (e), 0.852 ms (f).

FIG.17. Schematic illustration of hydrophobic repulsion by tail to initiate slippage

FIG.18. Dynamics of slippage mechanism of engulfing water layer over the drop. Snapshots obtained from high speed movies showing the process for t = 0 ms (a), 0.284 ms (b), 0.568 ms (c).

FIG.19. Dynamics of column formation. Snapshots obtained from high speed movies showing the process for t = 0 ms (a), 0.426 ms (b), 0.852 ms (c), 1.136ms.

FIG.20. Dynamics of column thinning. Snapshots obtained from high speed movies showing the process for t = 0 ms (a), 0.284 ms (b), 0.568 ms (c).

FIG.21. Schematic illustration of effect of Laplace pressure and Marangoni flow generated instability on drop neck thinning.

FIG.22. Morphological evolution of drop during thinning. Snapshots obtained from high speed movies showing the process for t = 0 ms (a), 0.71 ms (b), 0.781 ms (c), 1.207 ms (d), 1.562 ms (e), 1.917 ms (f), 2.343 ms (g), 2.556 ms (h), 2.698 ms (i).

FIG.23. Partial coalescence of surfactant laden water drop within pool of water.

FIG.24. Successive generation of secondary droplets for surfactant drop (CMC) (a) Primary drop; (b) Secondary drop; (c) Tertiary drop; (d) Quaternary drop; (e) Pentanary drop.

FIG.25. Effect of concentration on individual coalescence time.

**List of Tables**

Table 1: Diameter and physical properties of individual drop for different concentration

Table 2: Non-dimensional numbers related to drop physical properties Nomenclature

CMC: Critical Micelle Concentration

$P_{out}$ : Outside pressure of drop

$P_{in}$ : Inside pressure of drop

$\gamma$ : Static Surface Tension of the liquid contained in drop

R: Radius of drop

**Supplementary Information (SI)**

Table 2: Non-dimensional numbers related to drop physical properties

Movie S1: Video for water drop impinging on pool of water; duration of video is 13.27 s

Movie S2: Video for surfactant laden water drop (CMC) impinging on pool of water; duration of video is 19.92 s

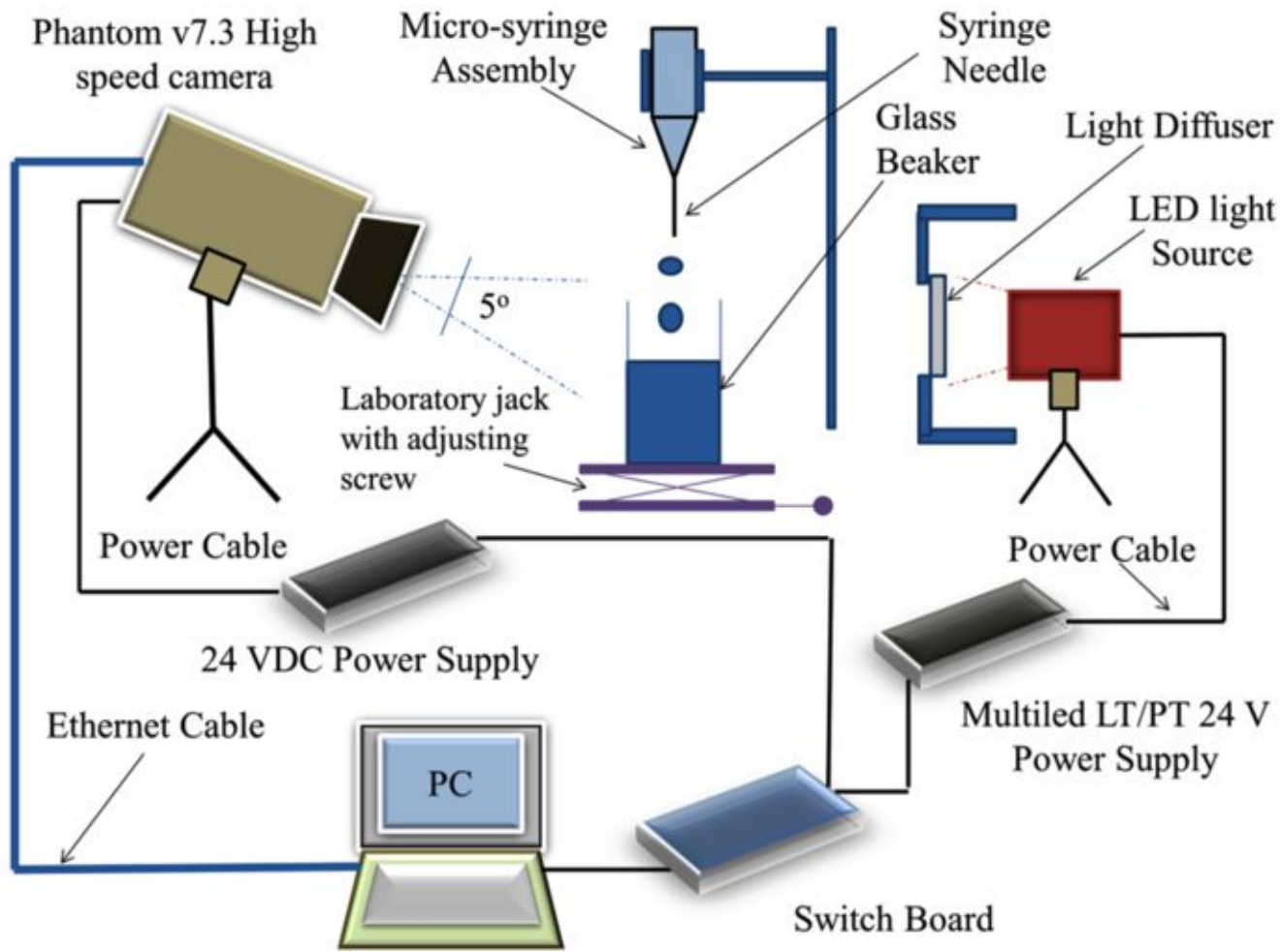

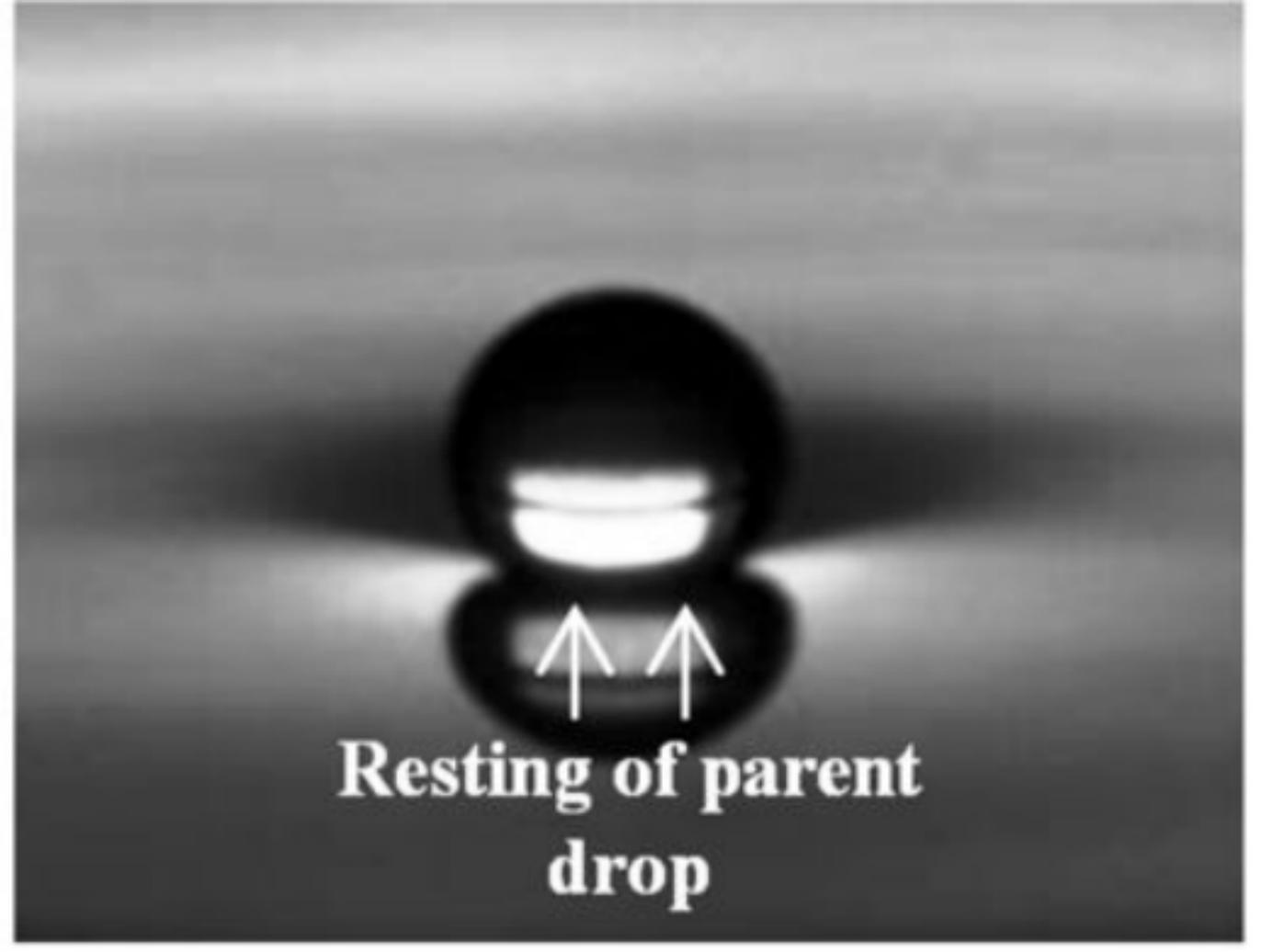

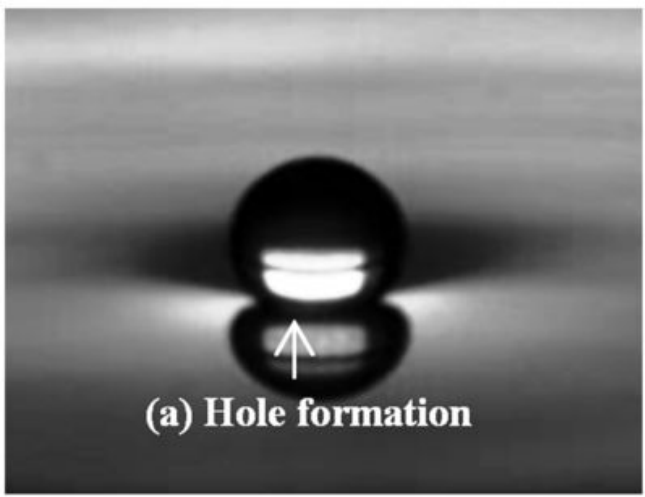
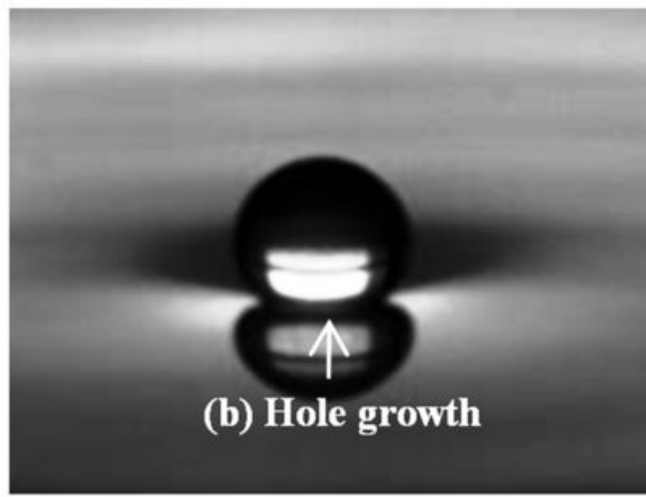
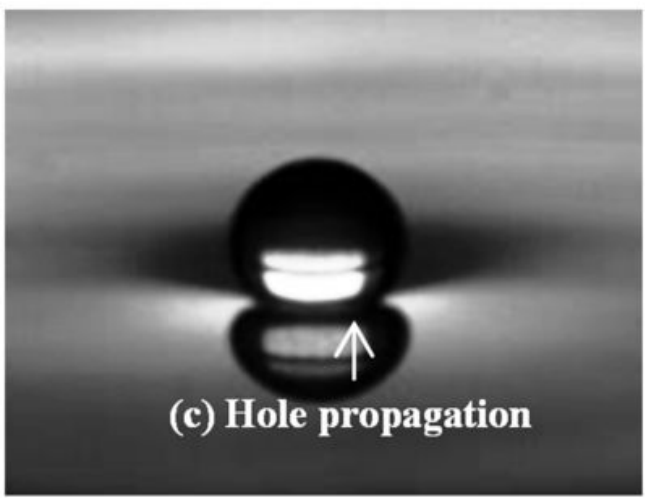
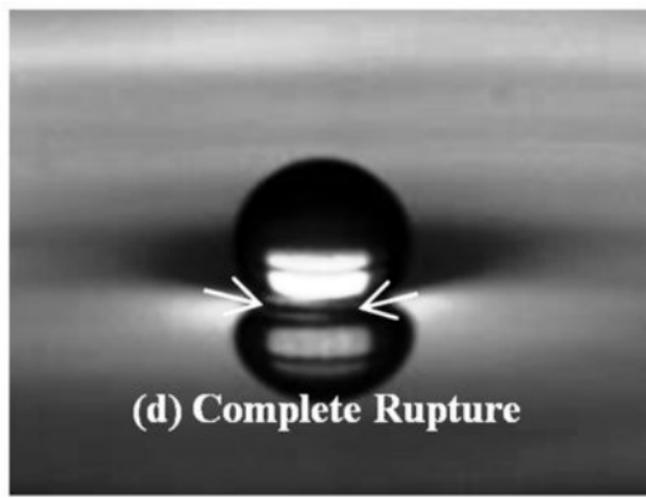

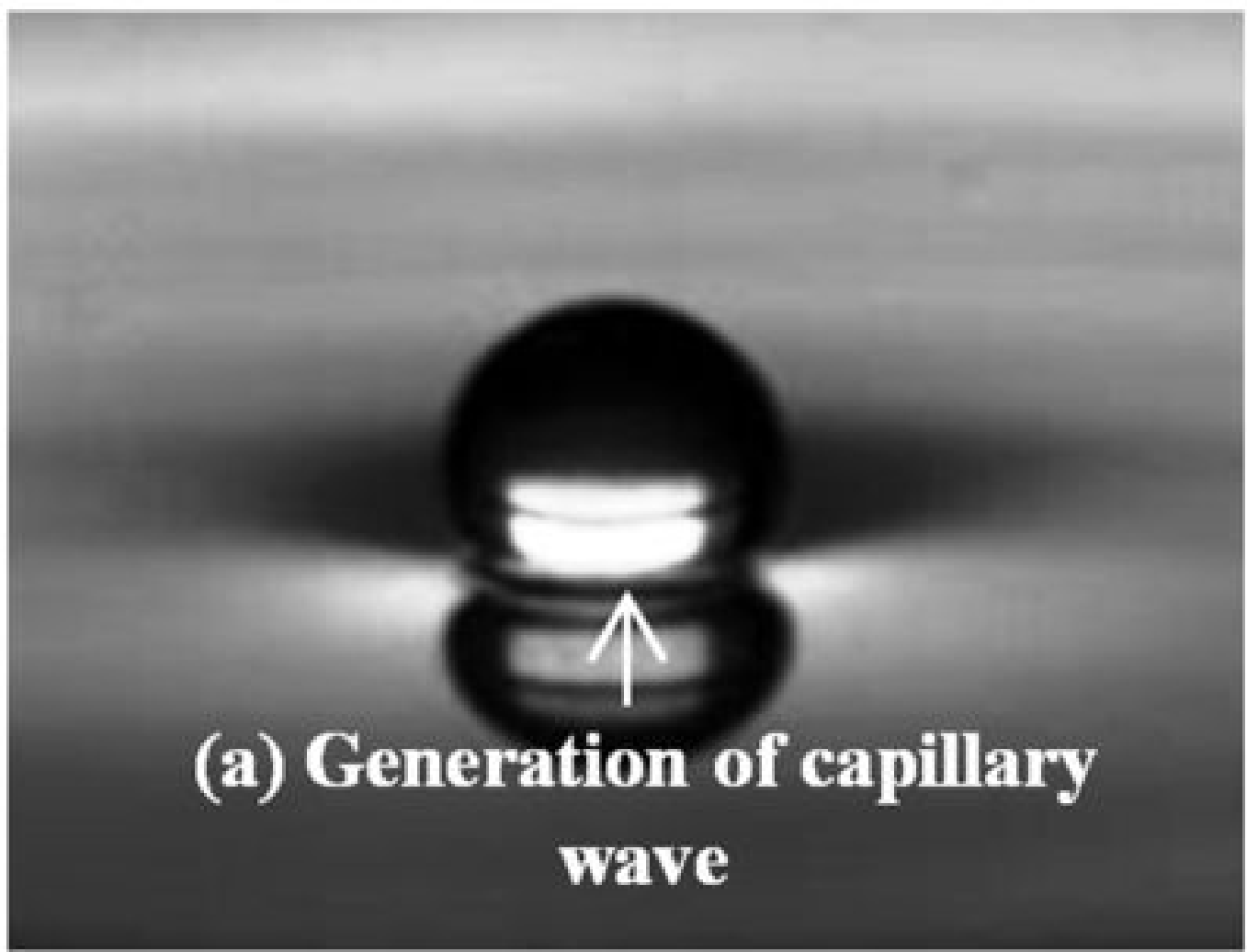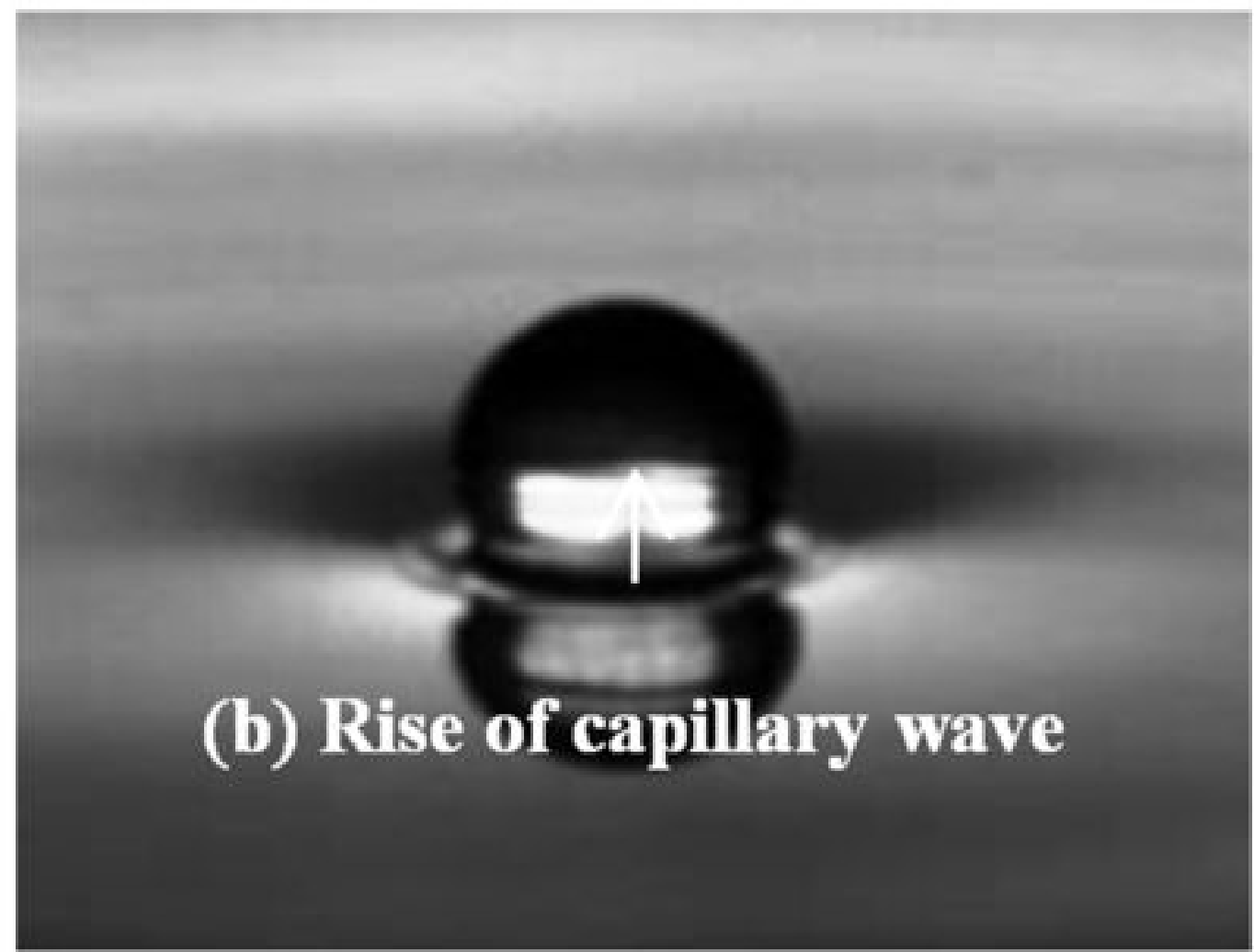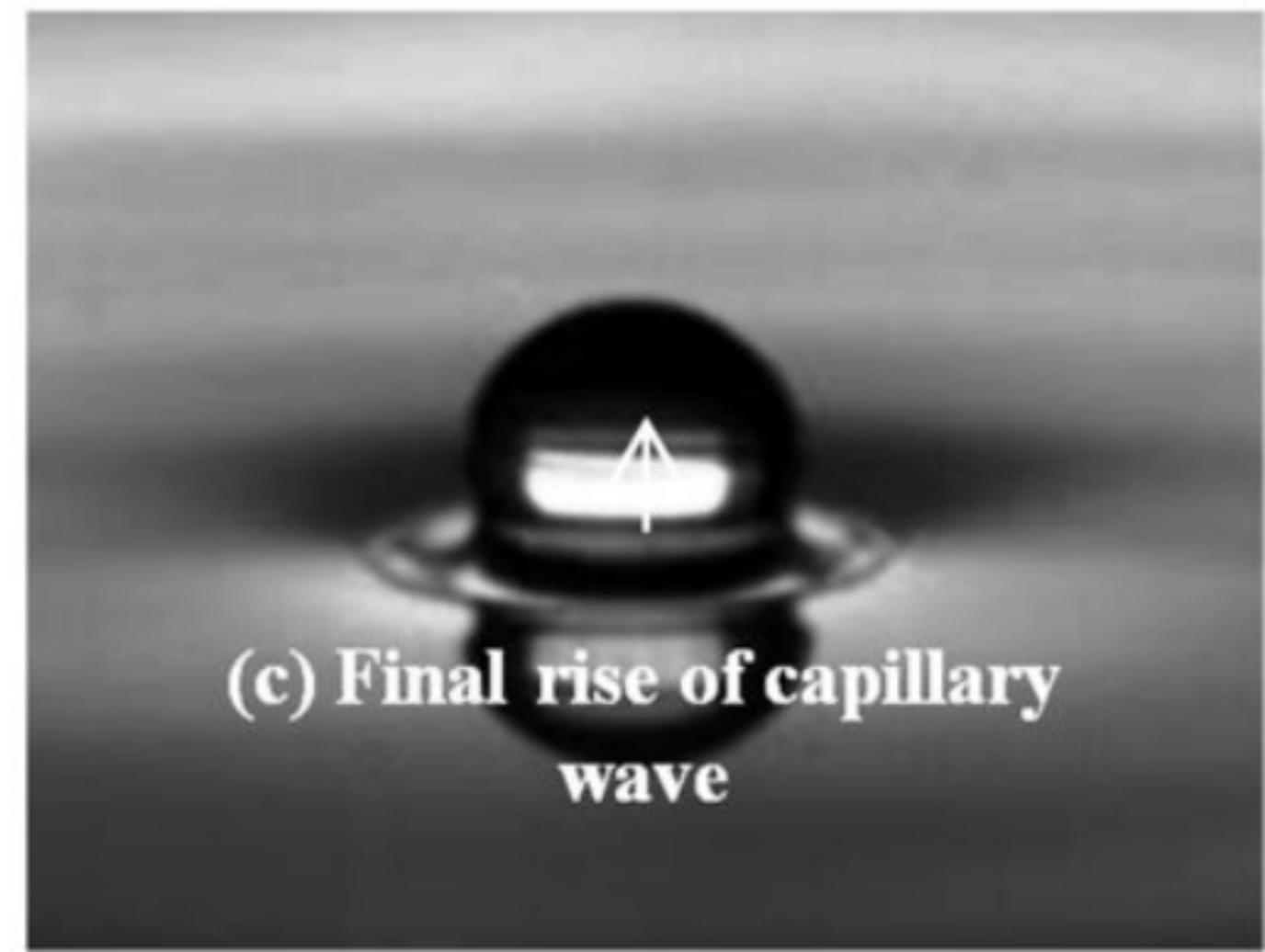

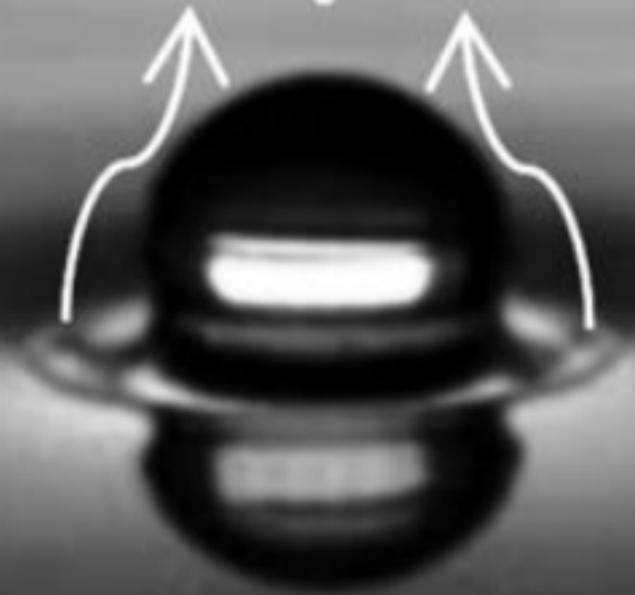 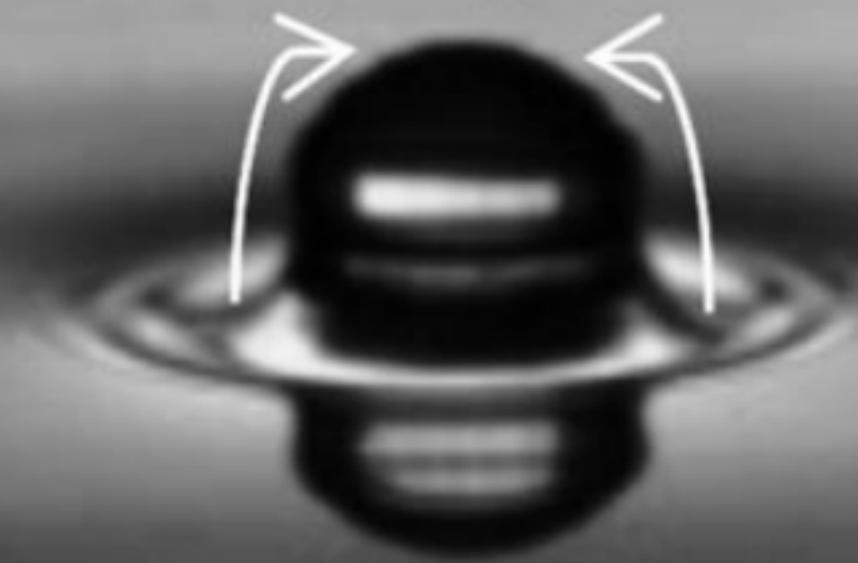 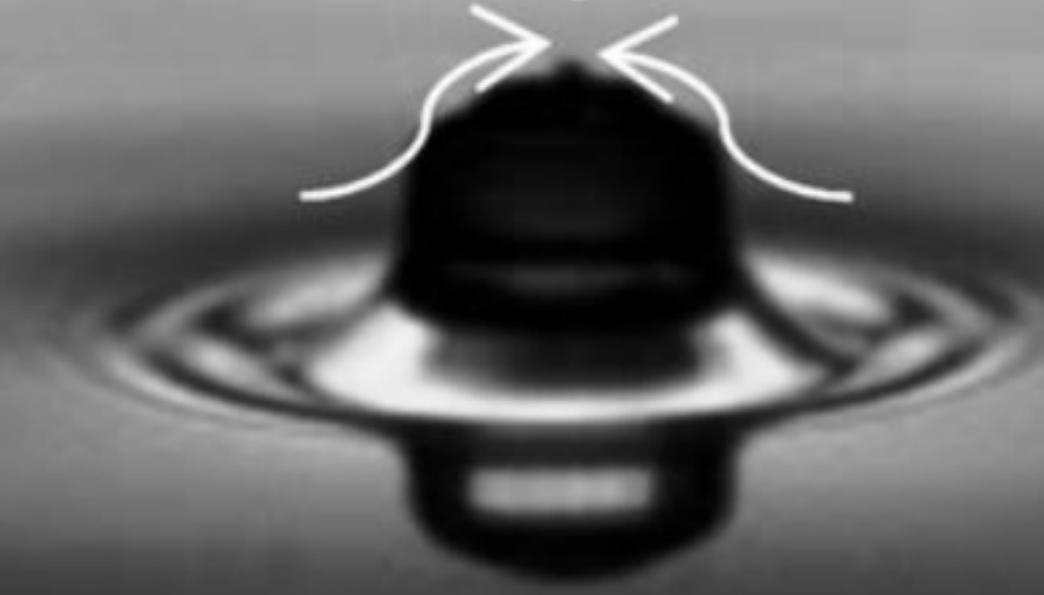

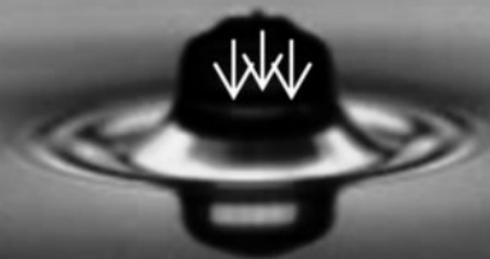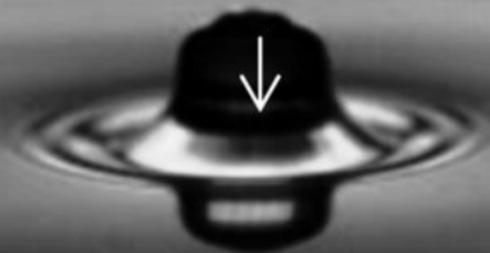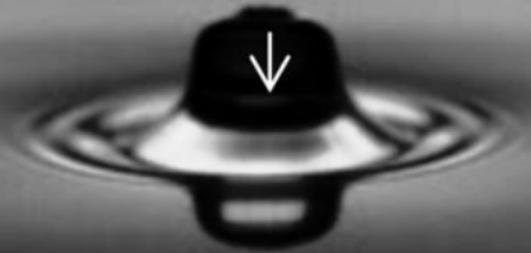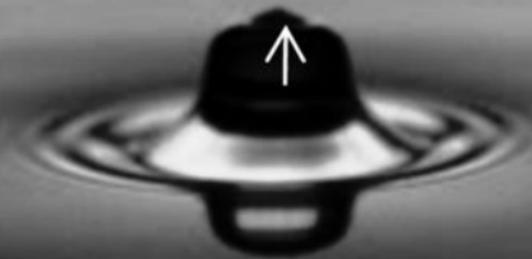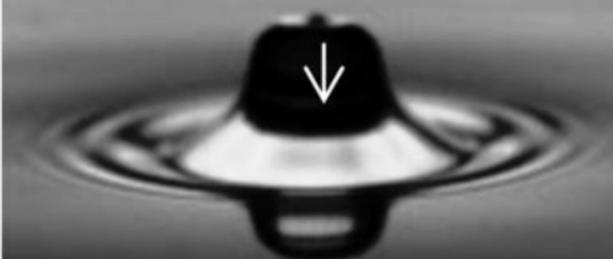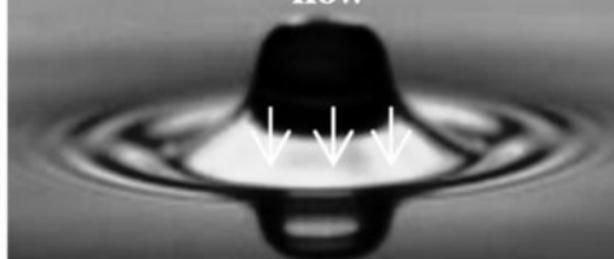

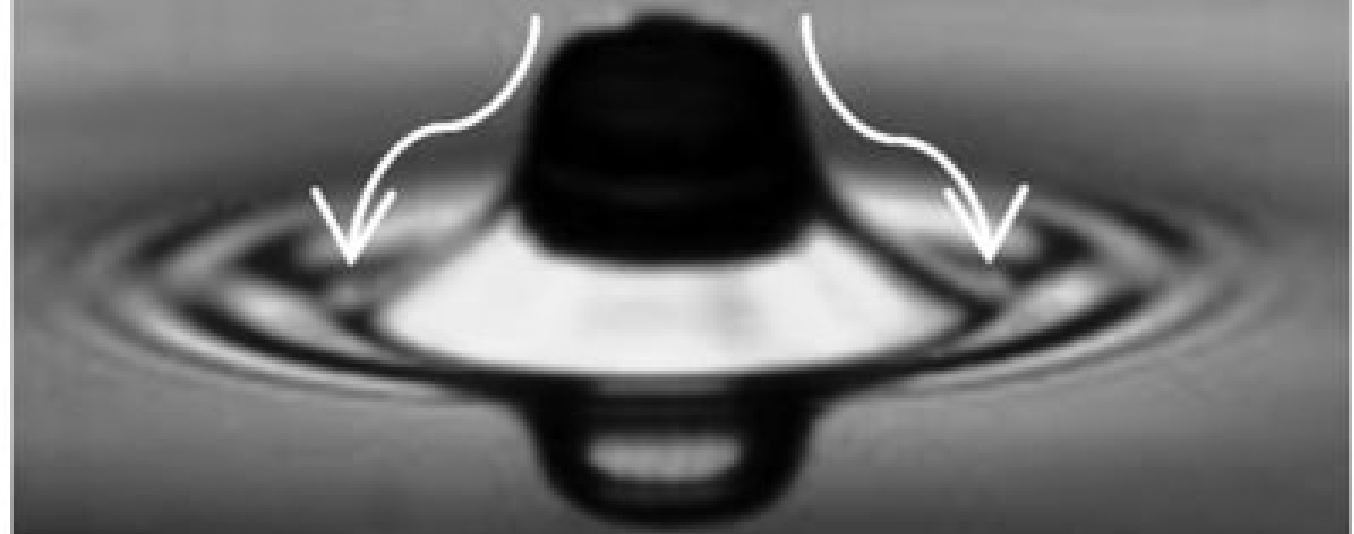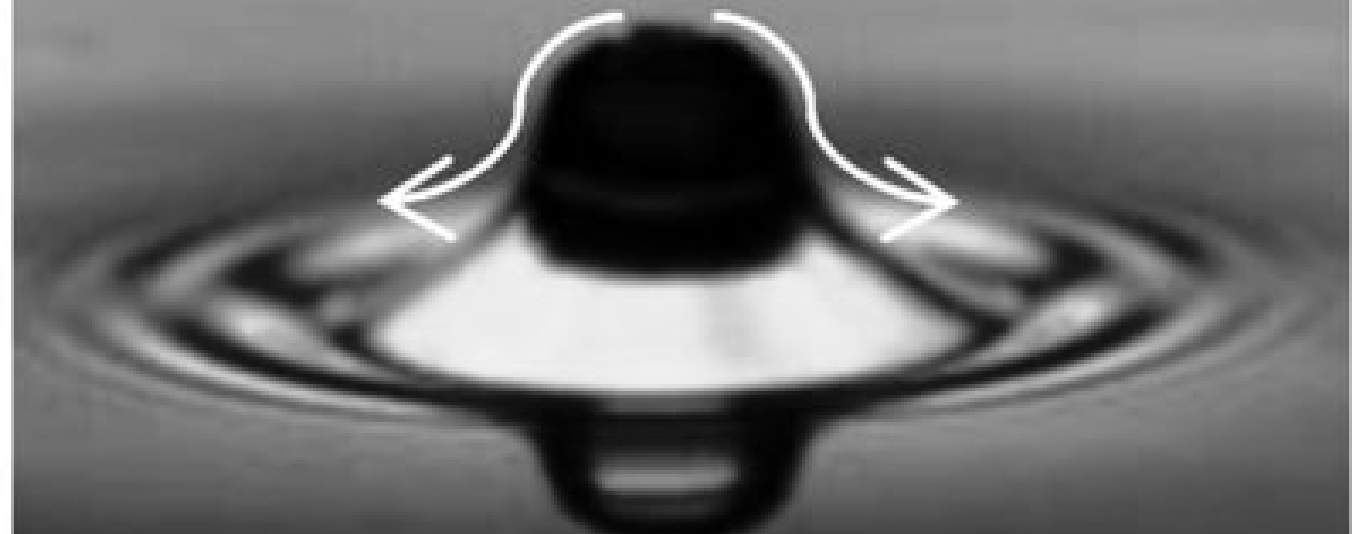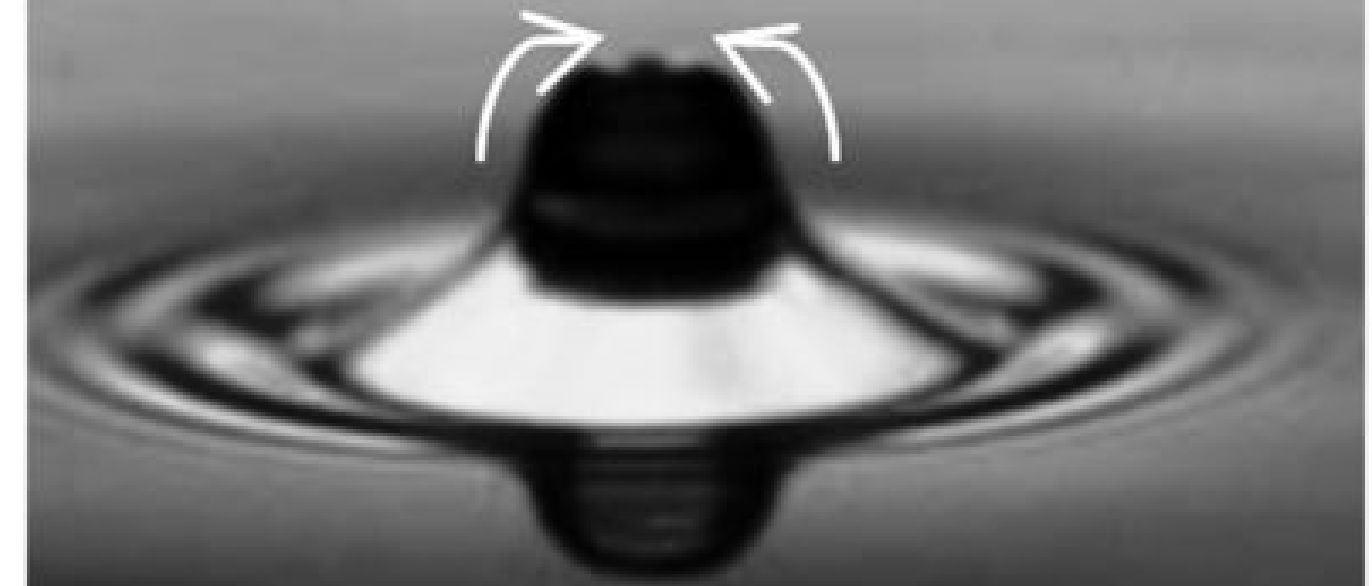

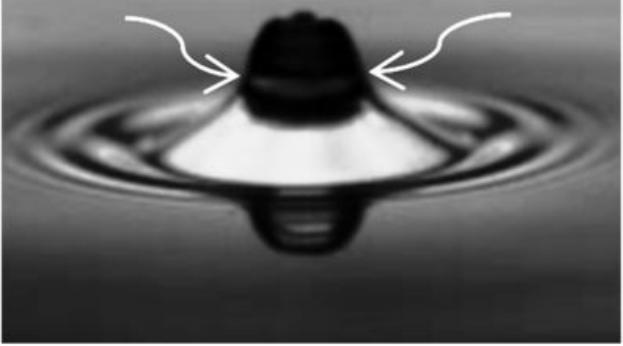 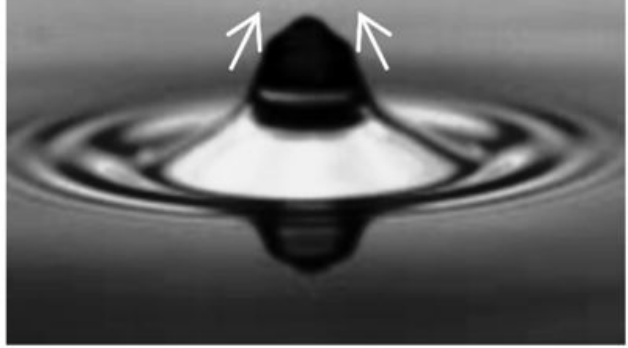 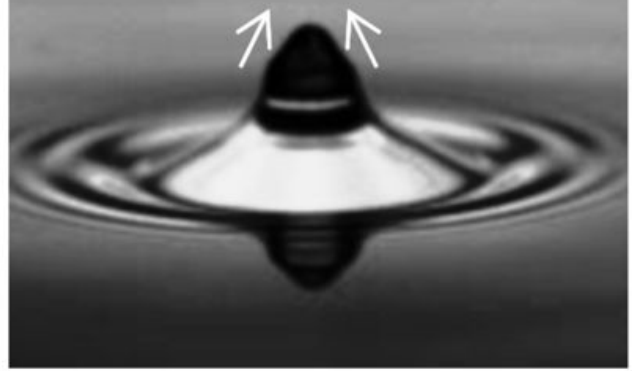
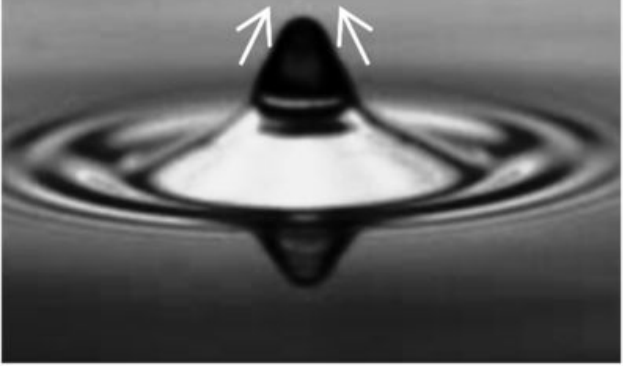 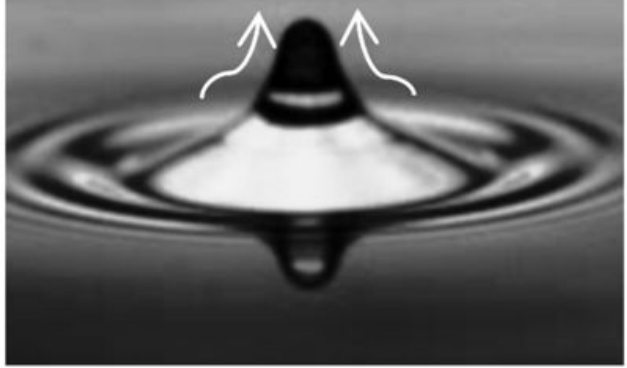 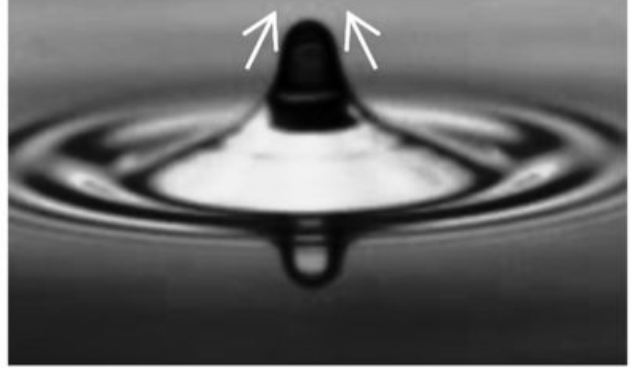

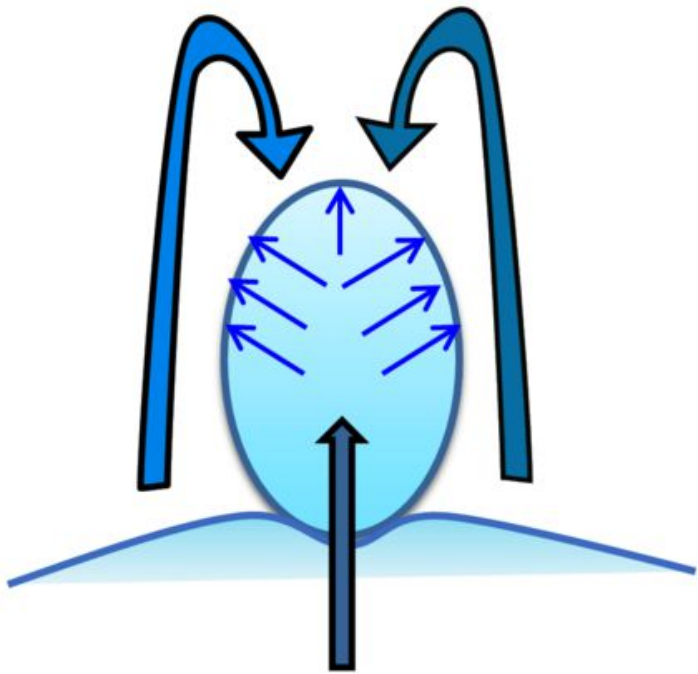 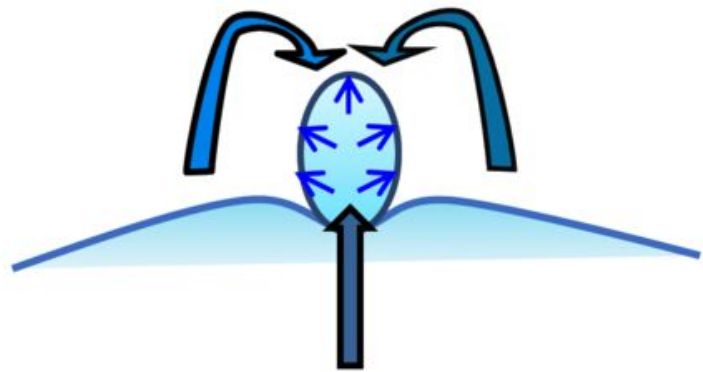

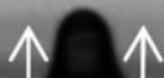
(a) Formation of pillar like drop

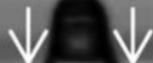
(b) Height reduction of pillar

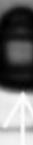
(c) Formation of pinned drop

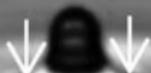
(d) Collapsing of pinned drop

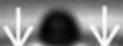
(e) Continuation of collapse

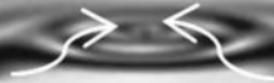
(f) Final Collapse

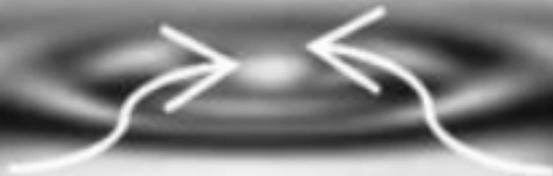

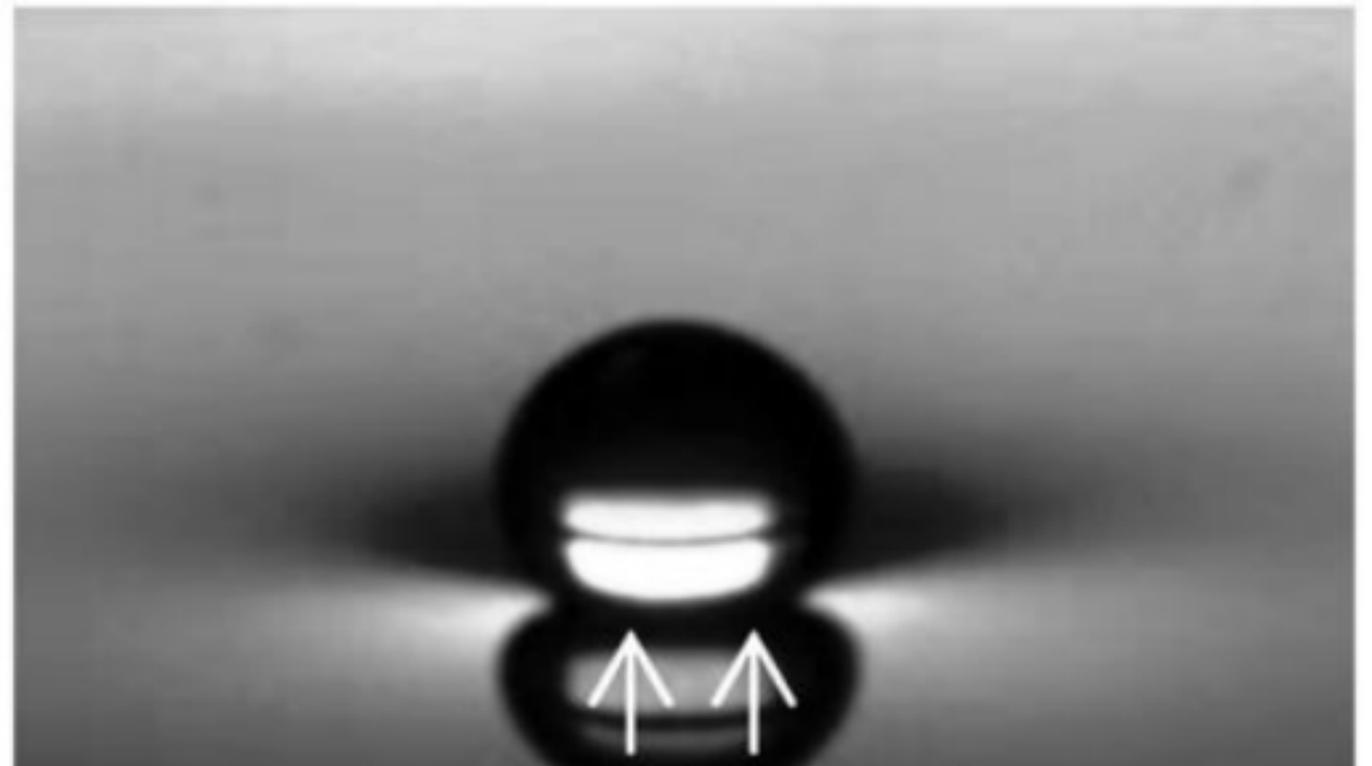

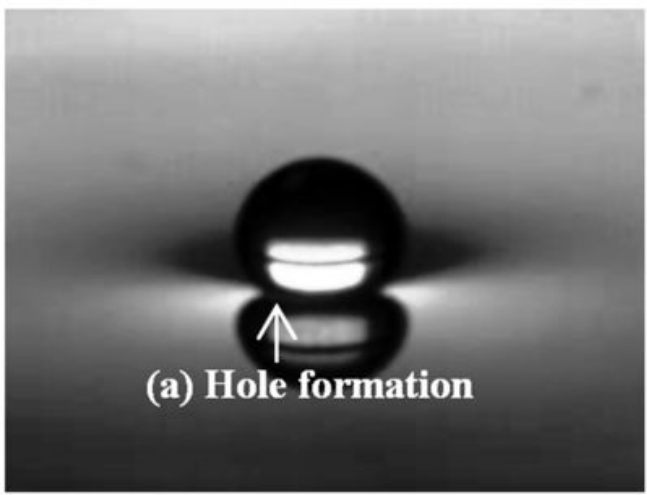 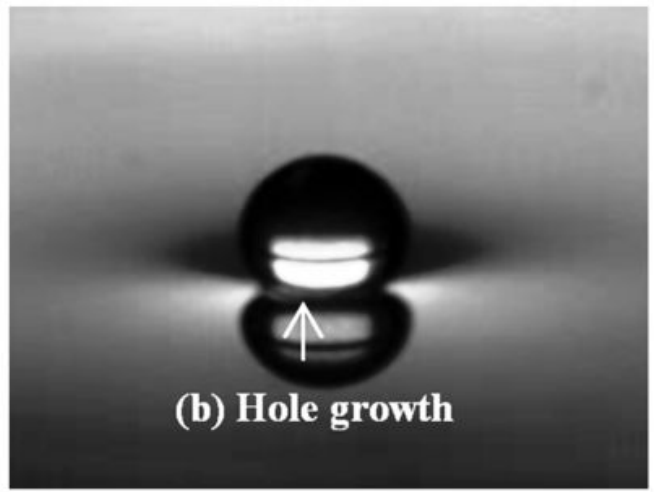 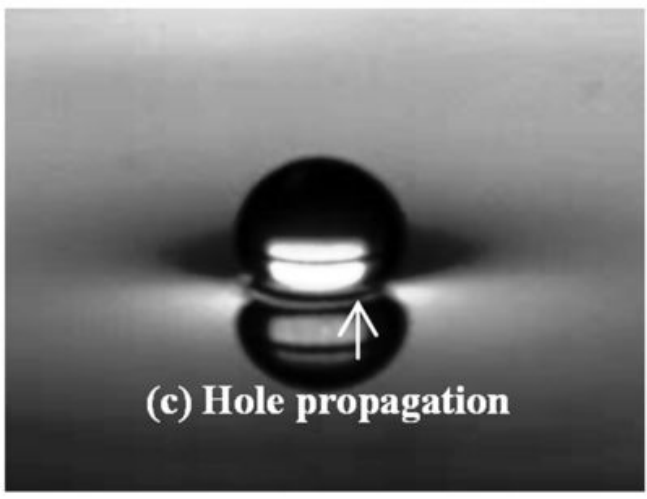 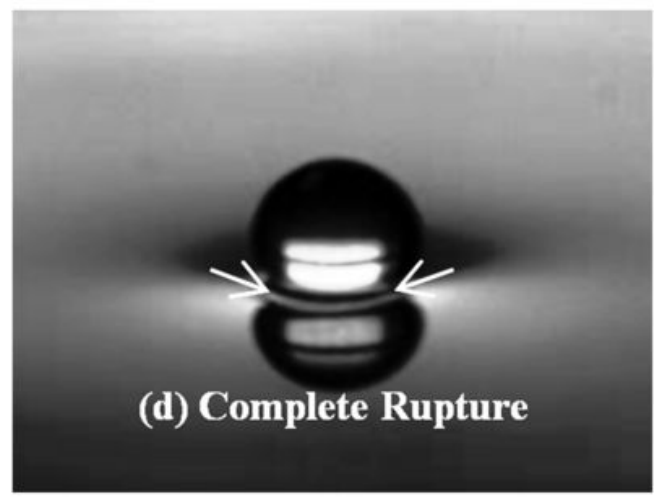

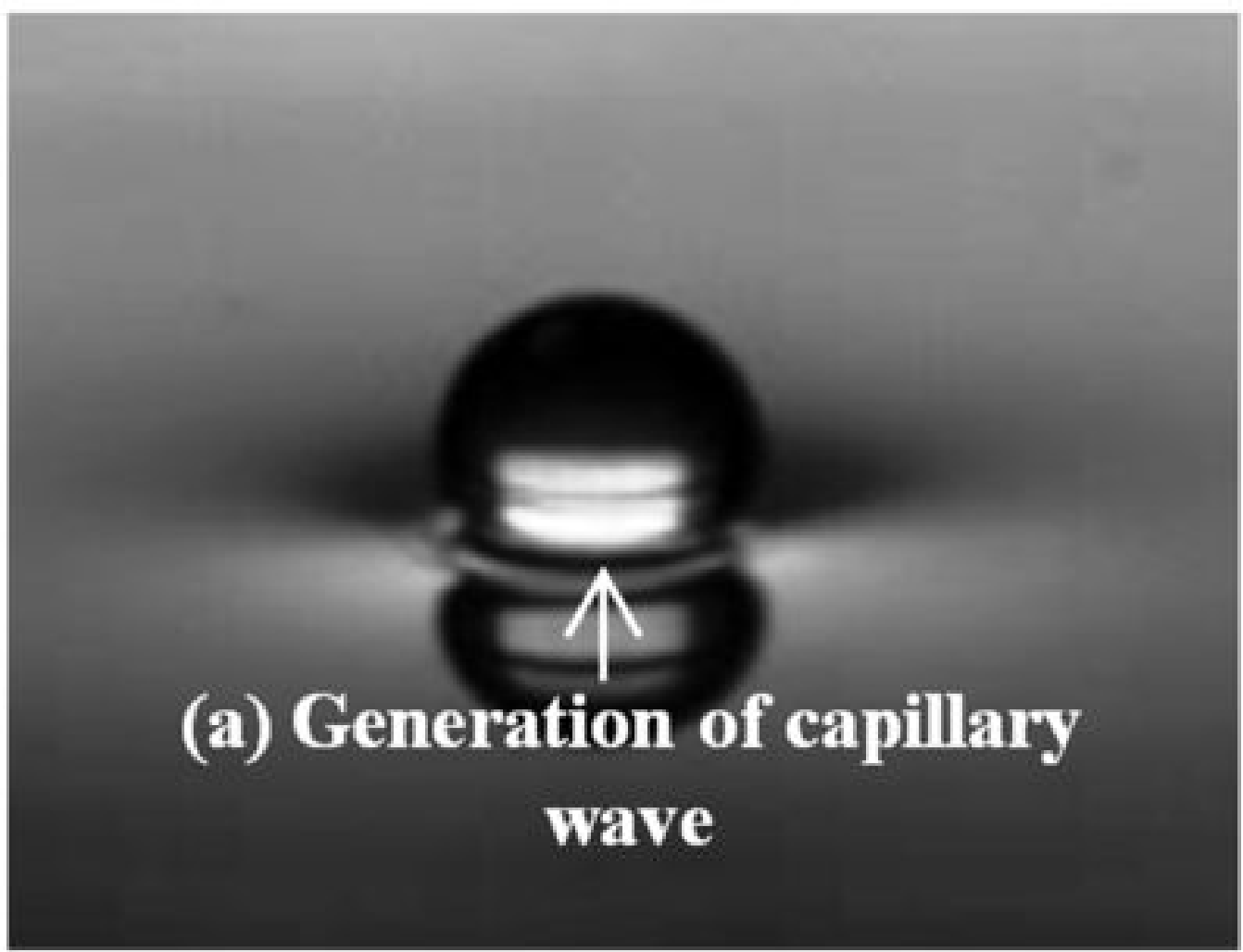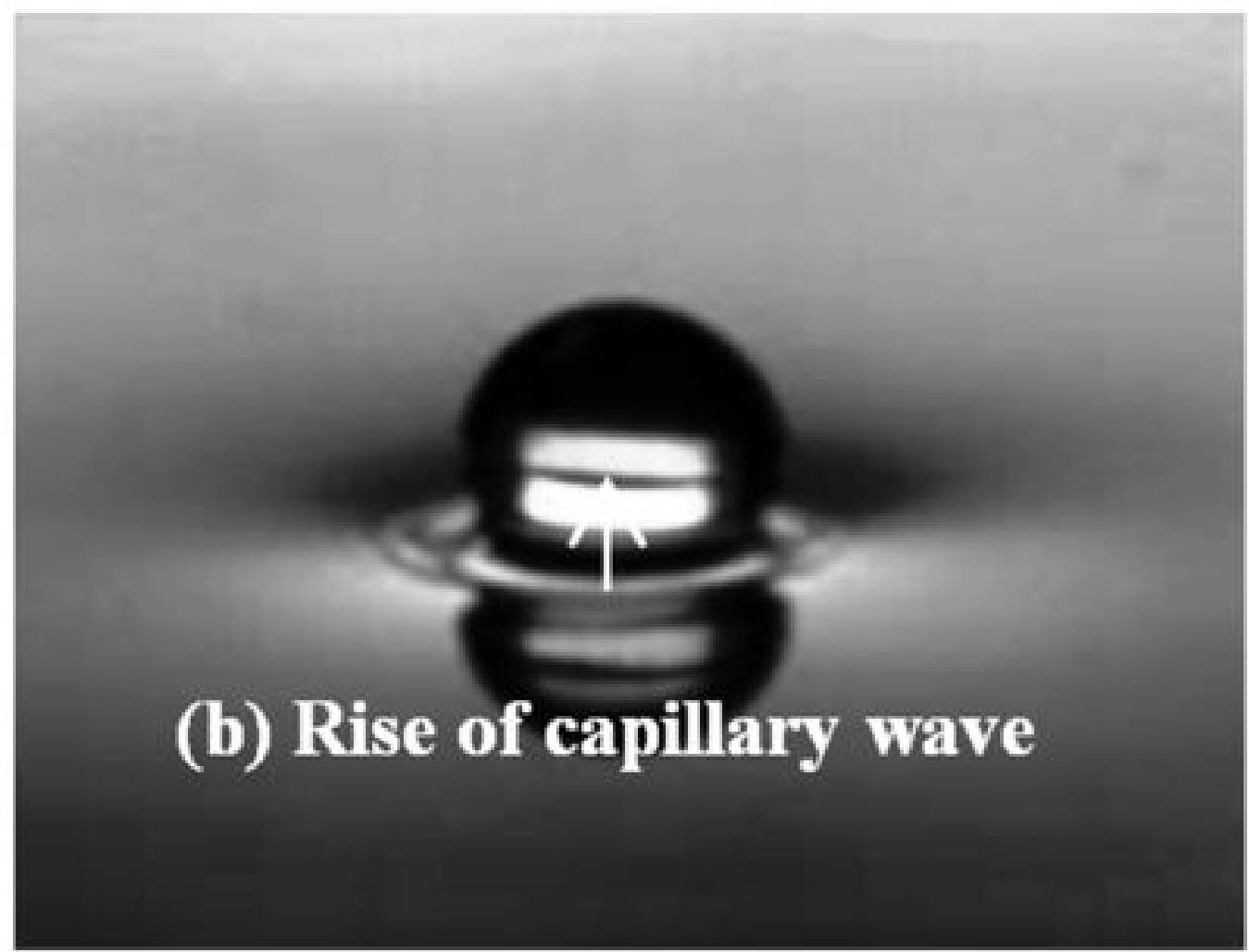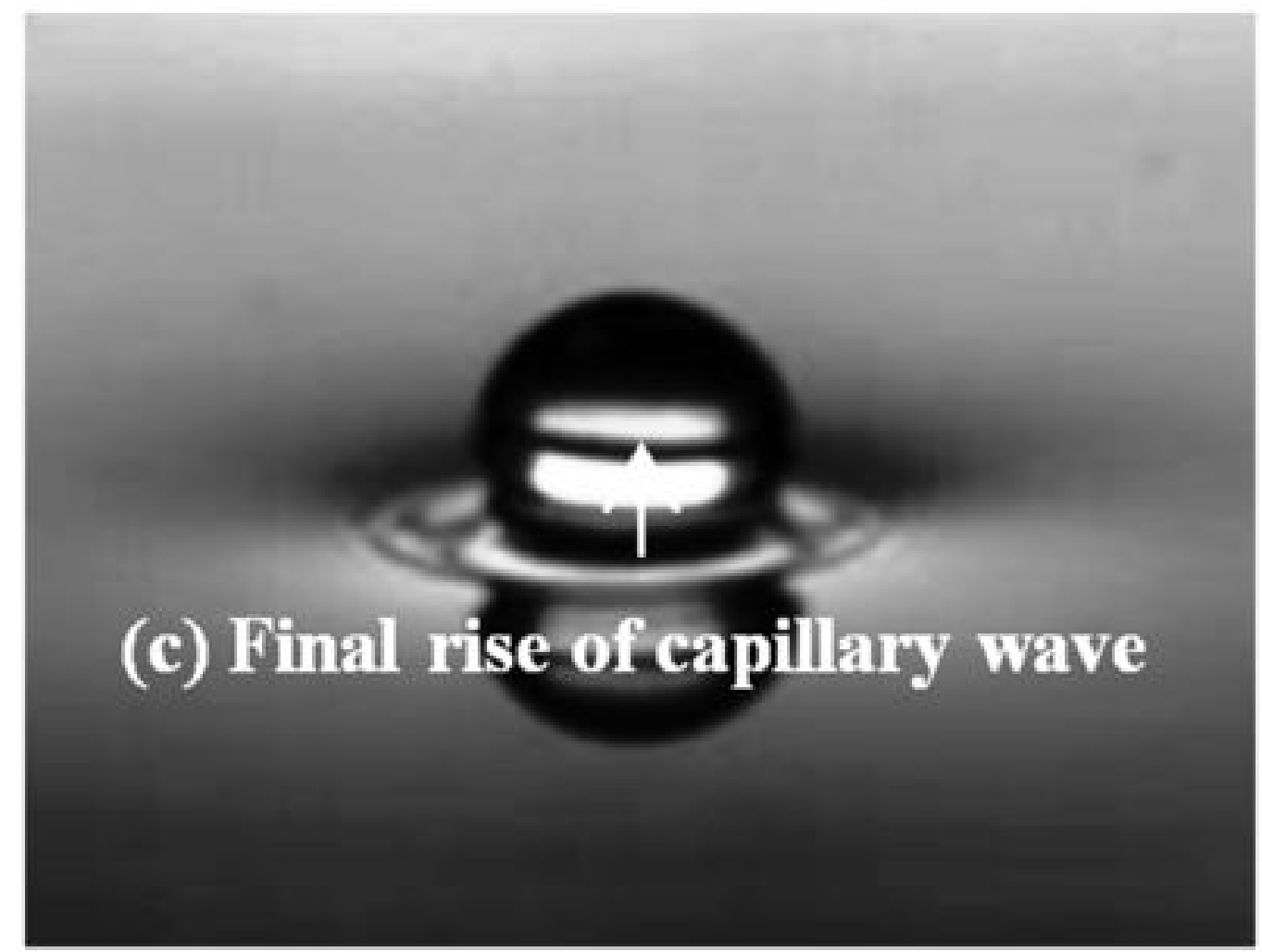

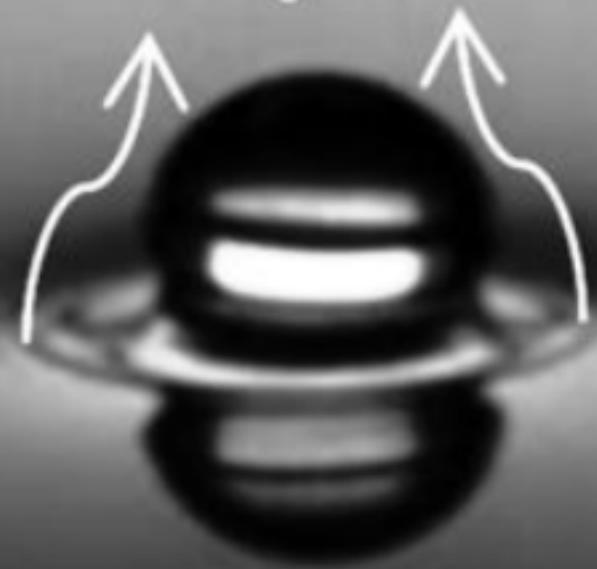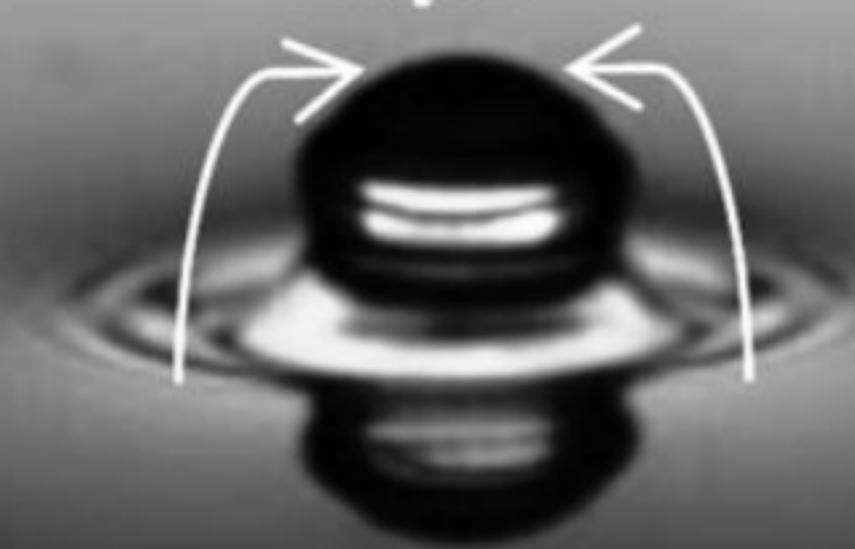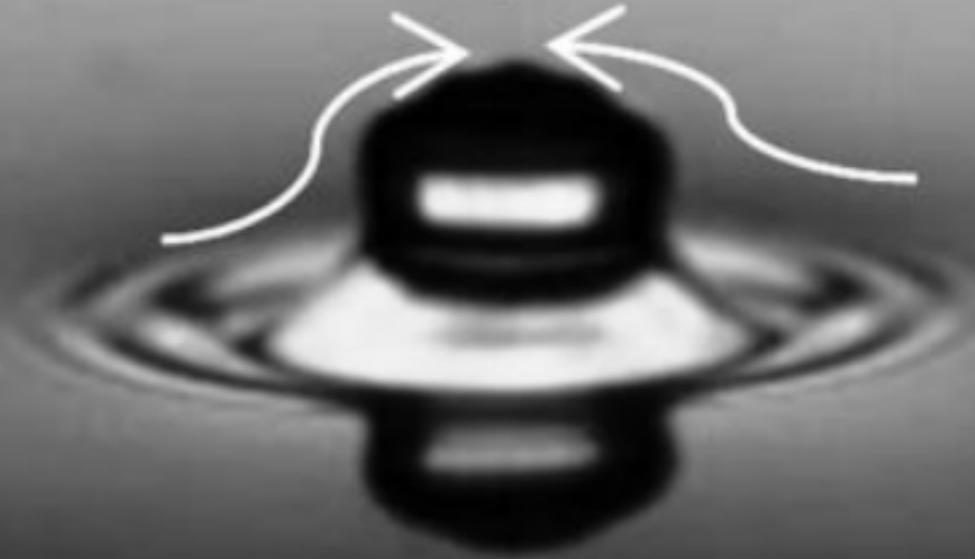

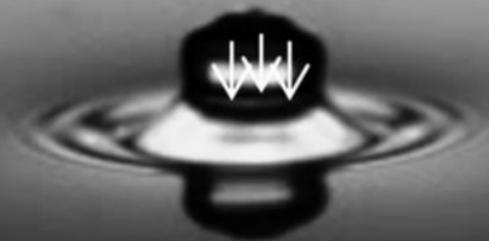 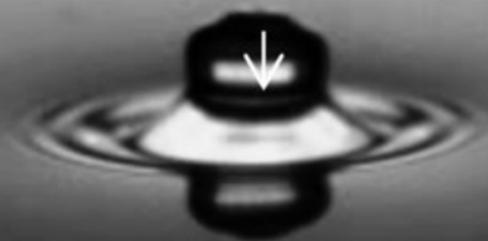 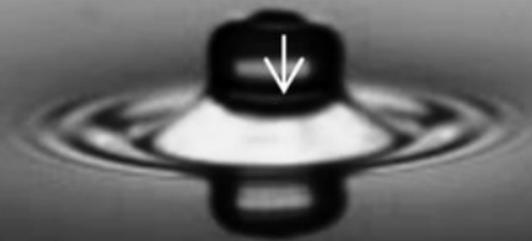
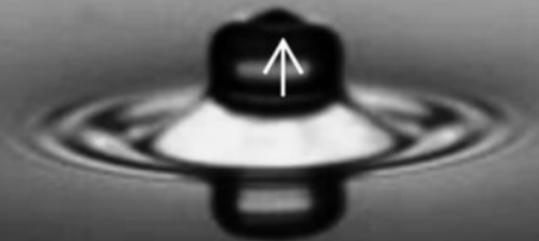 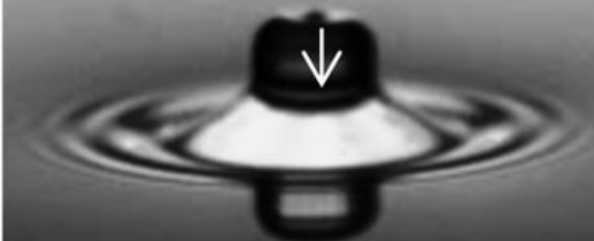 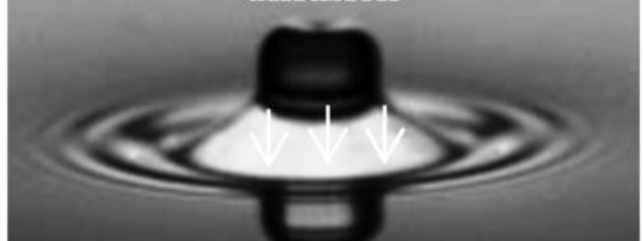

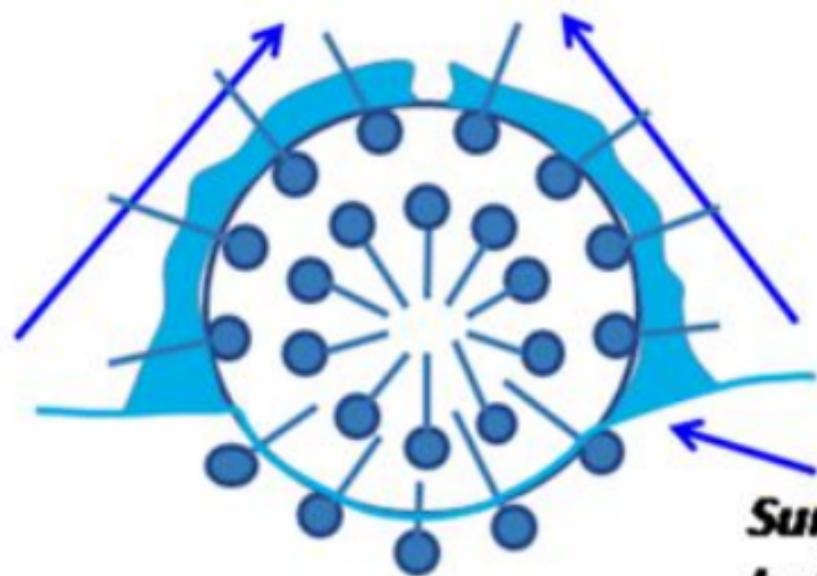 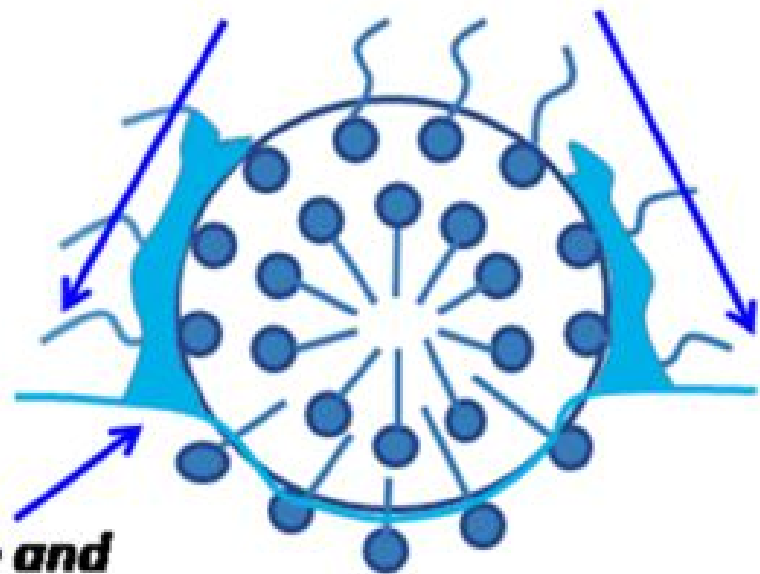

*Engulfing Water Layer*     *Slippage of Water Layer*

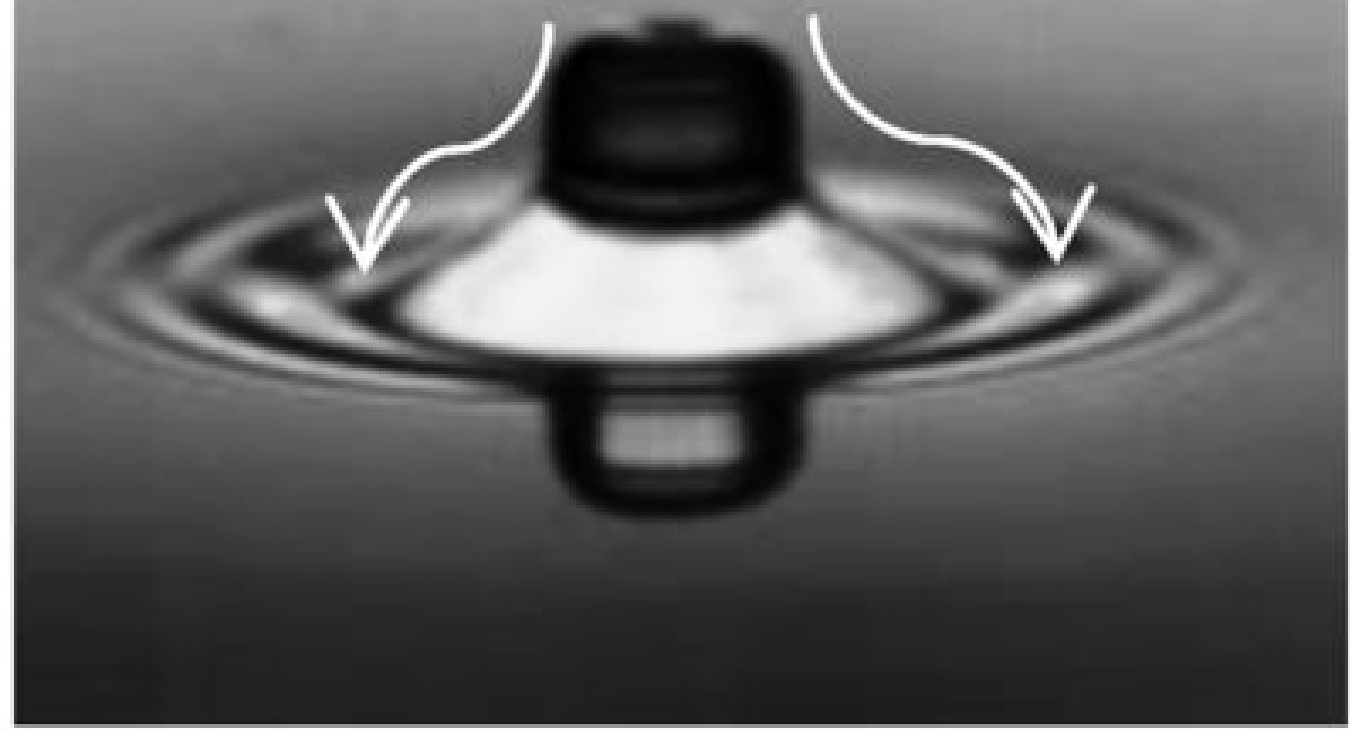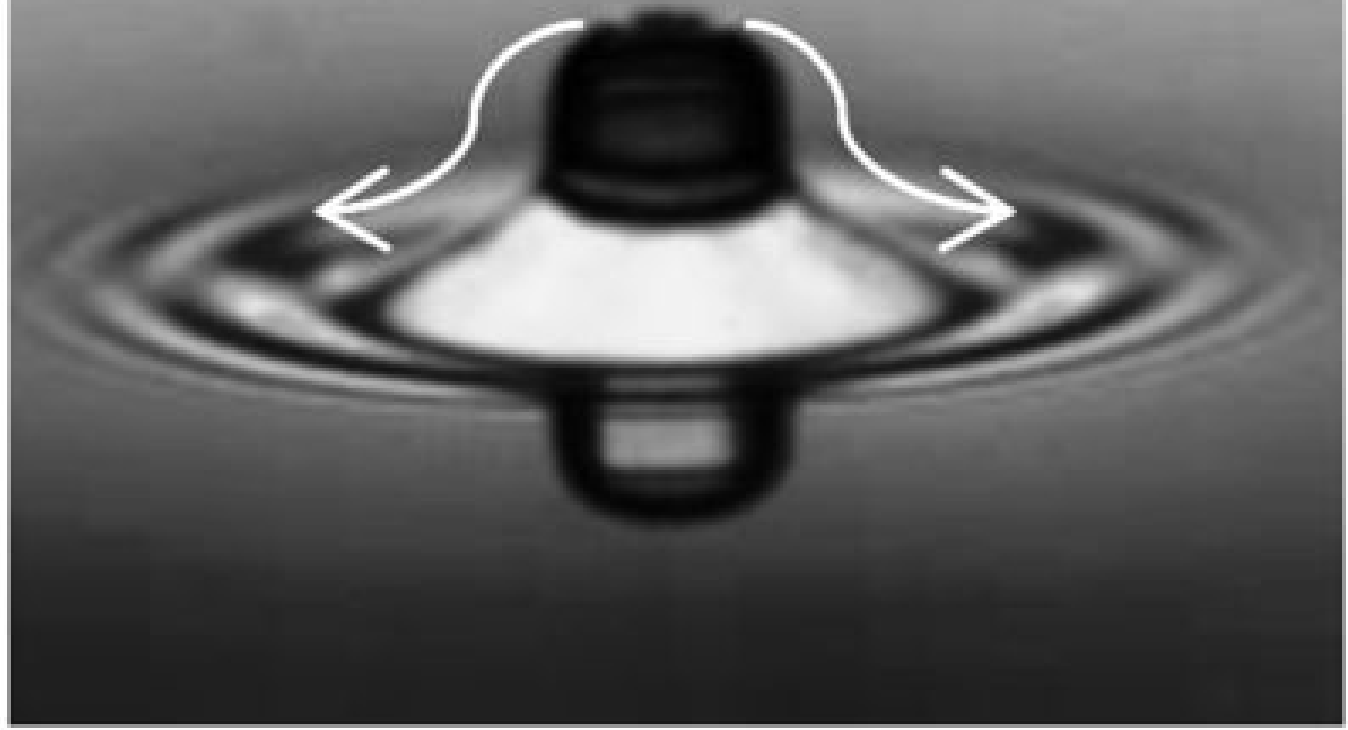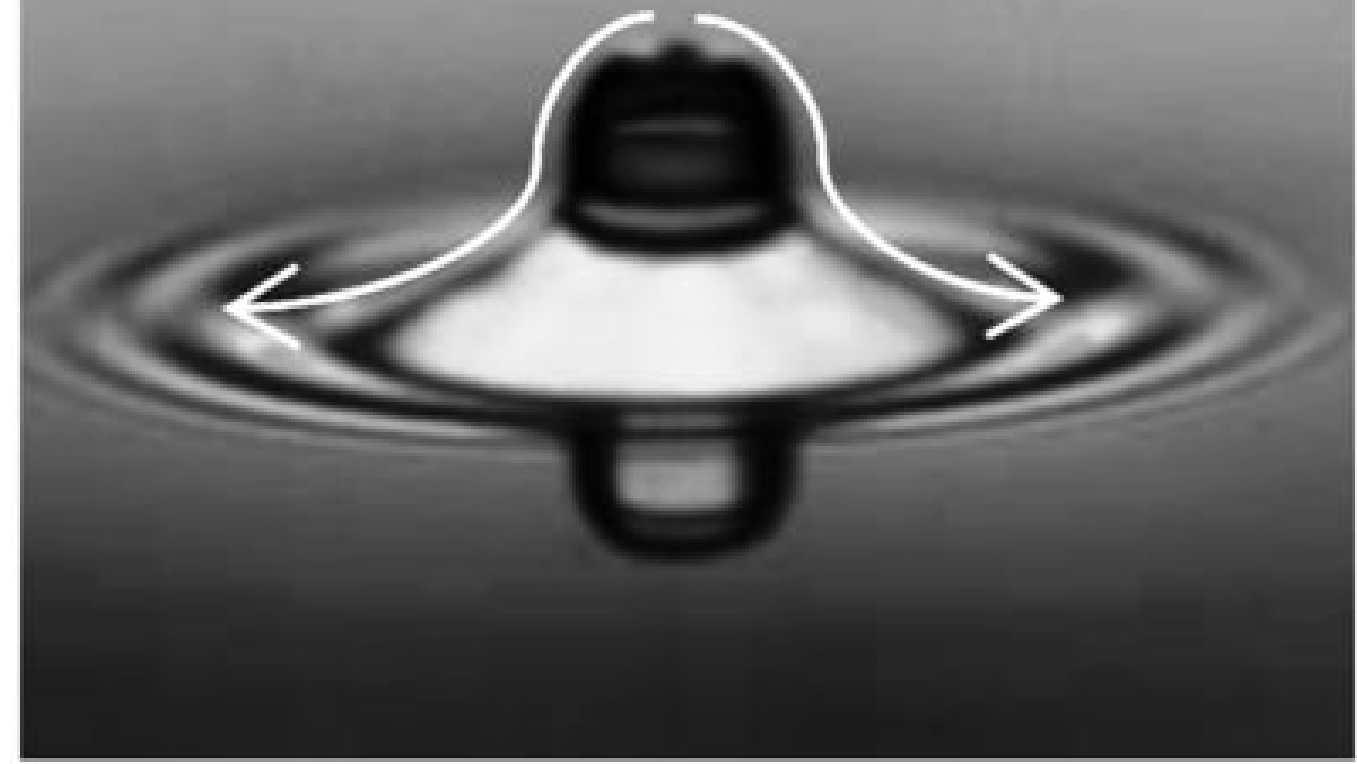

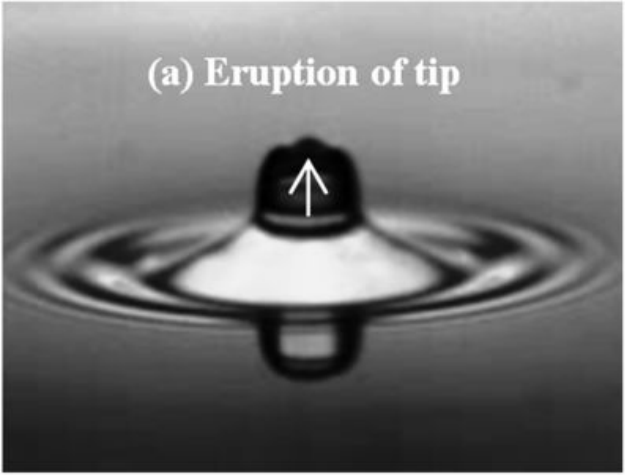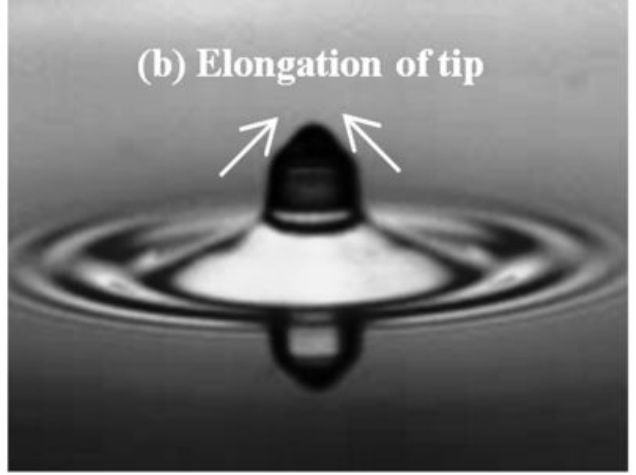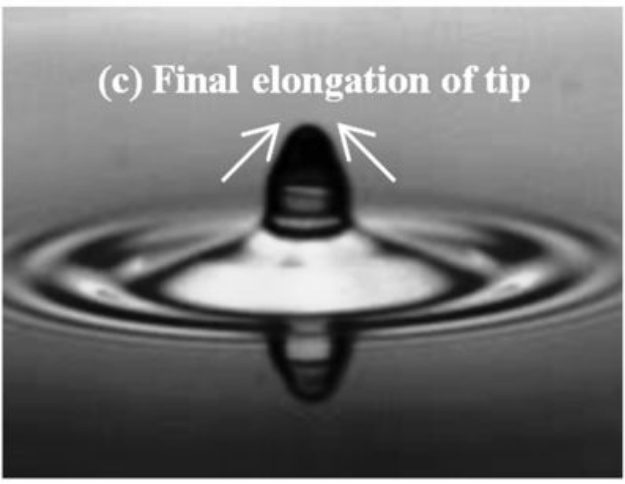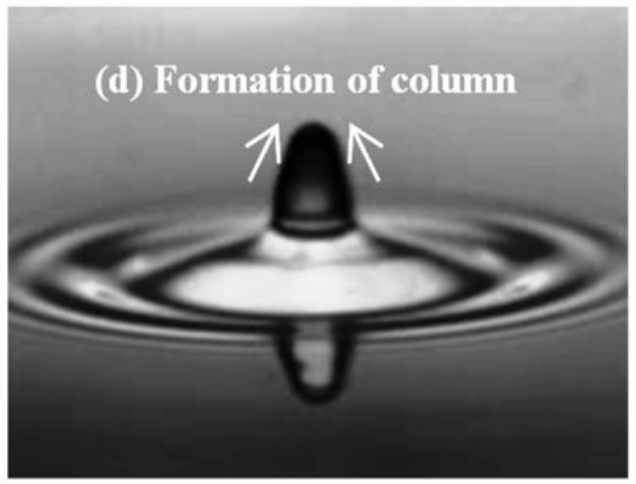

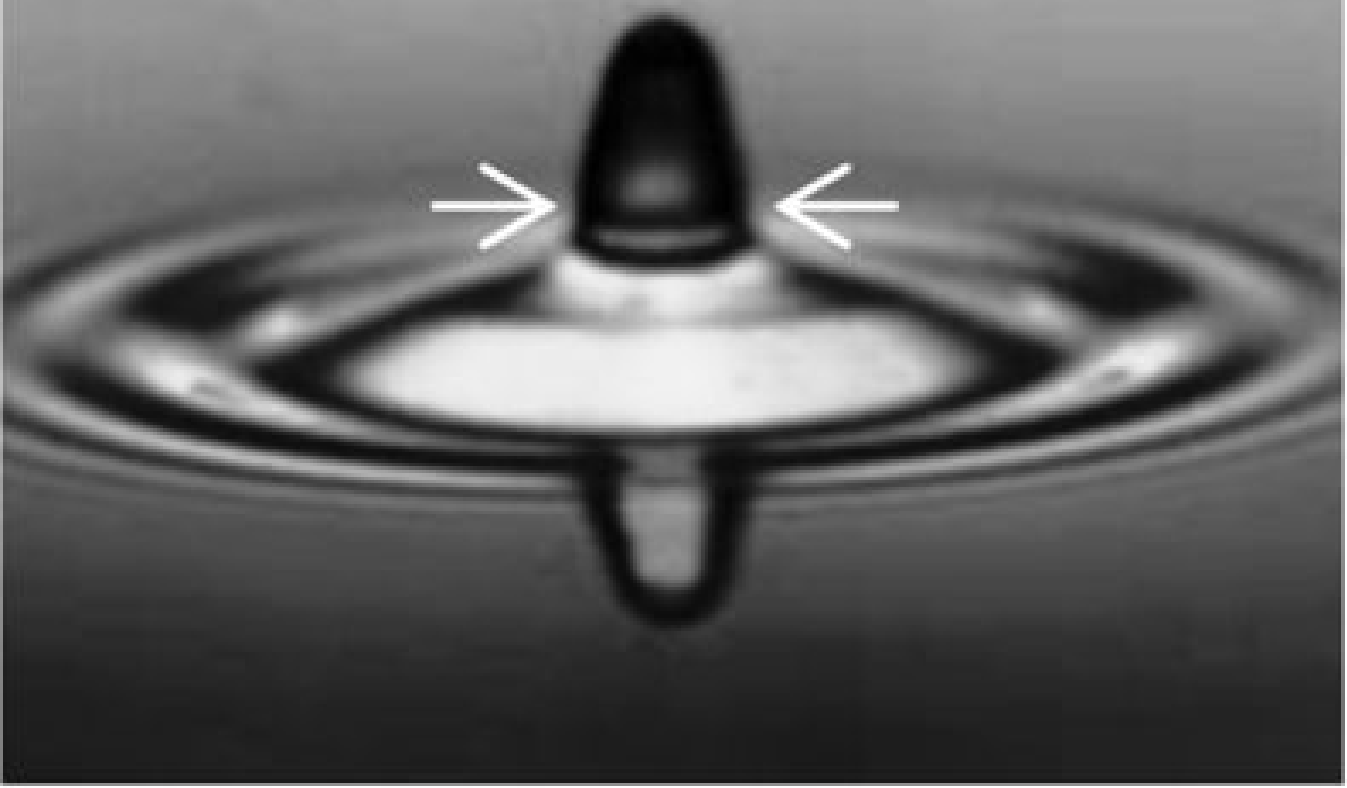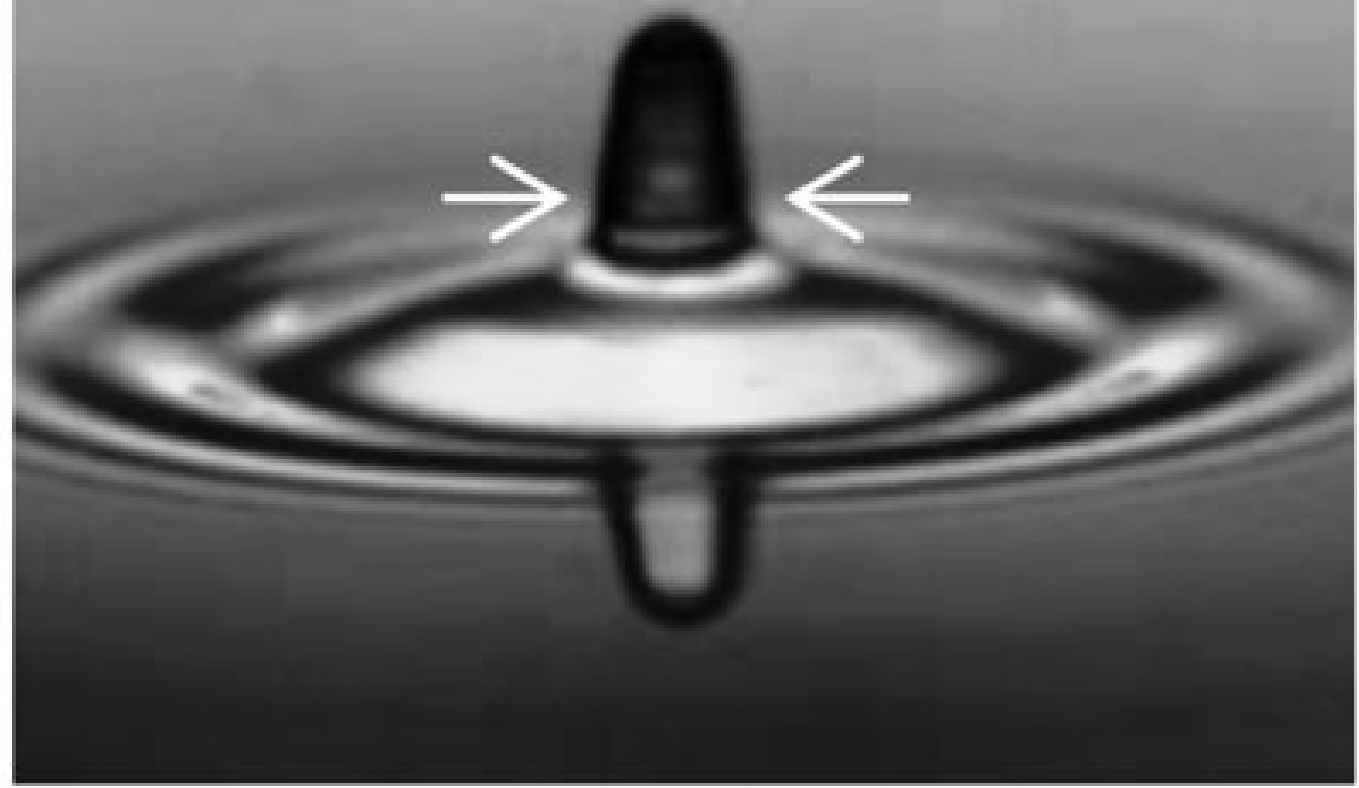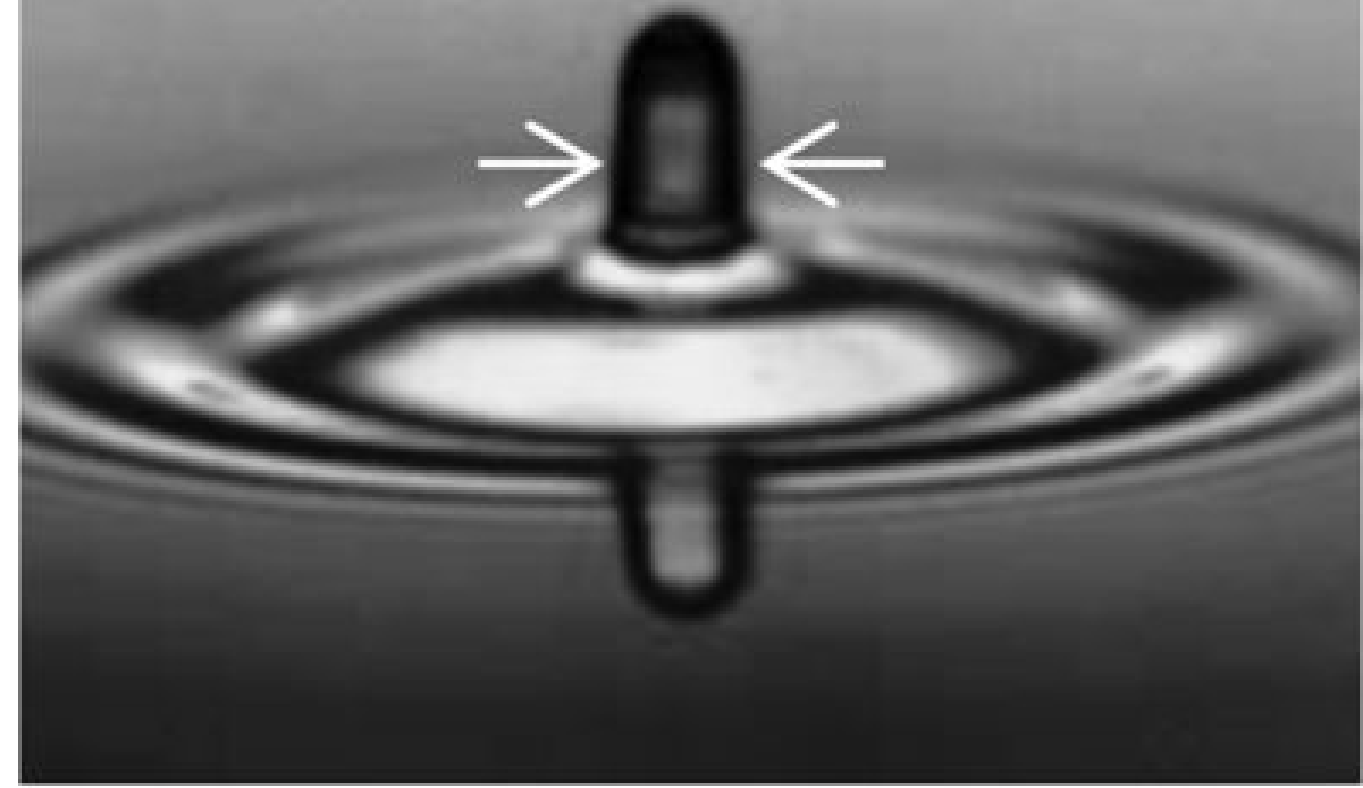

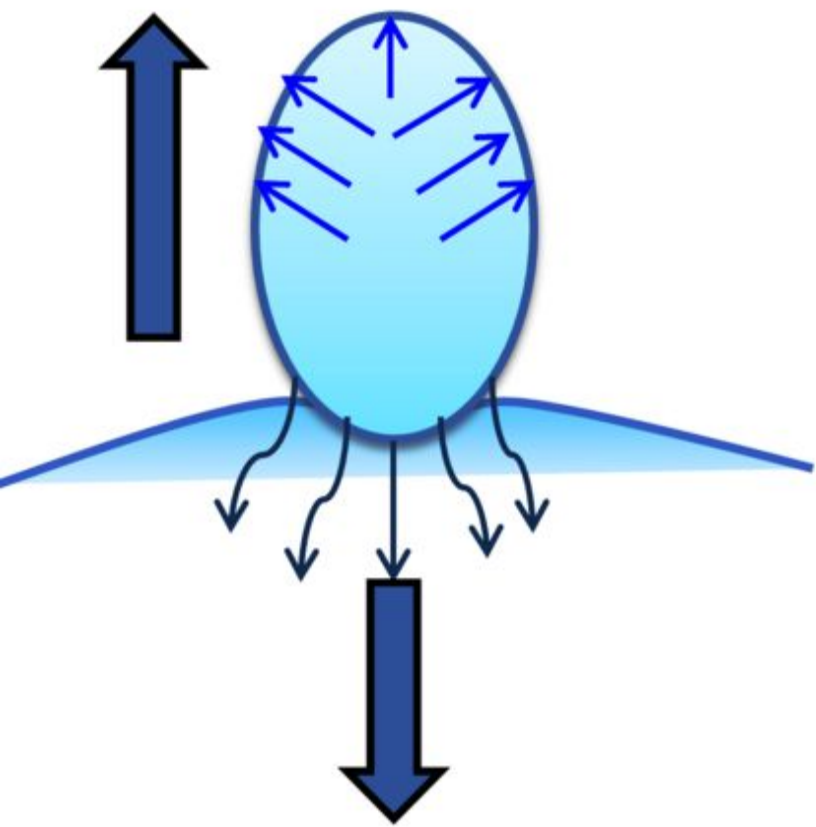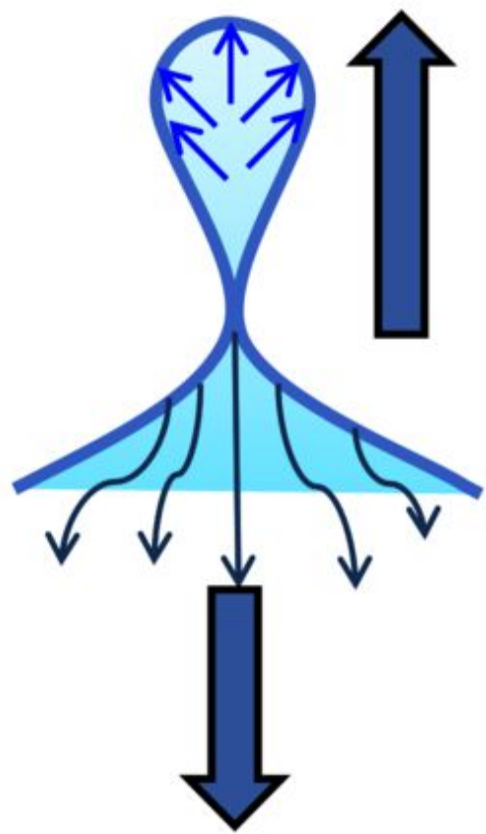

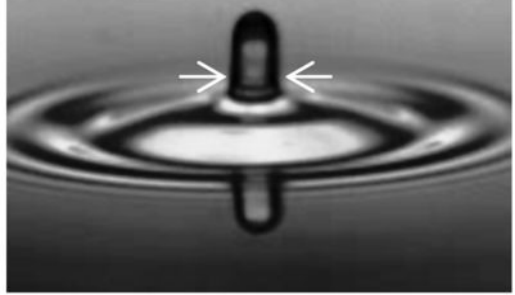 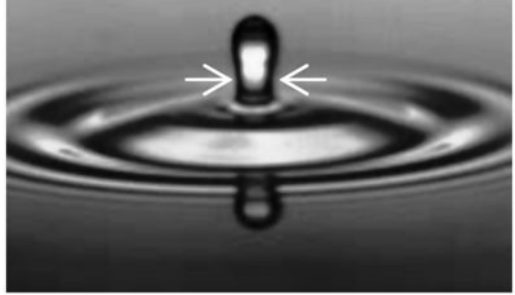 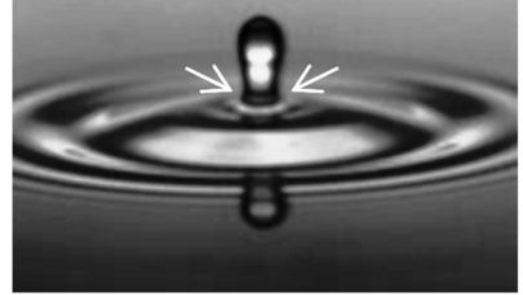
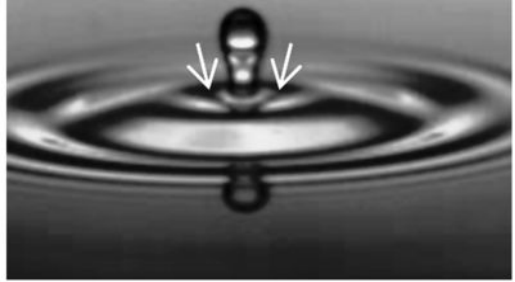 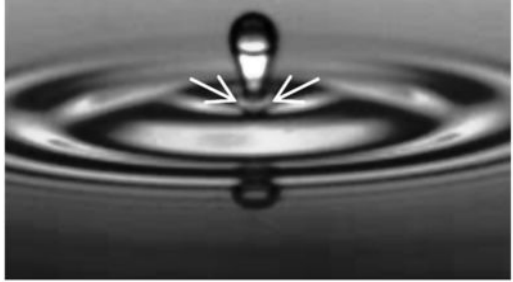 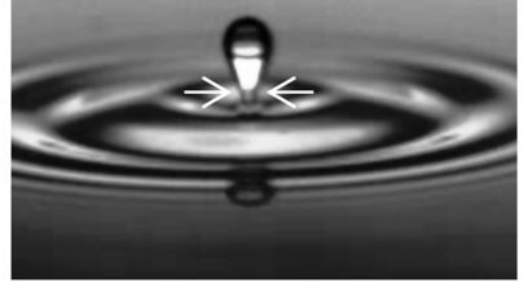
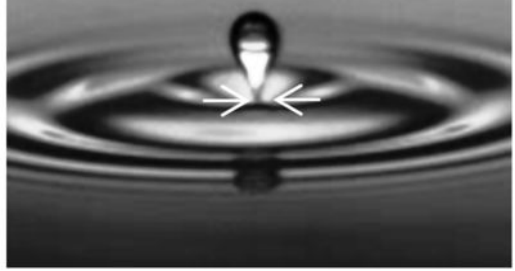 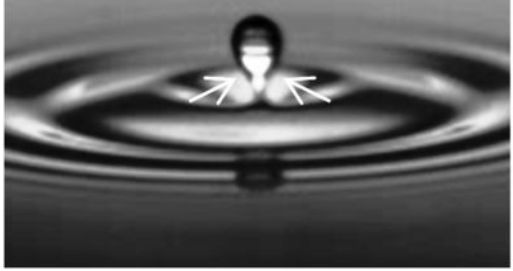 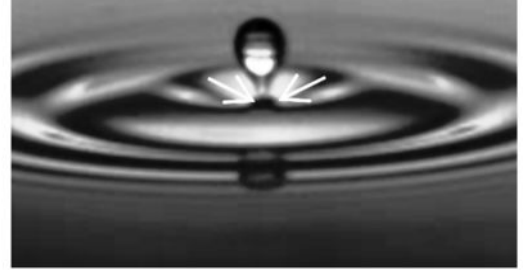

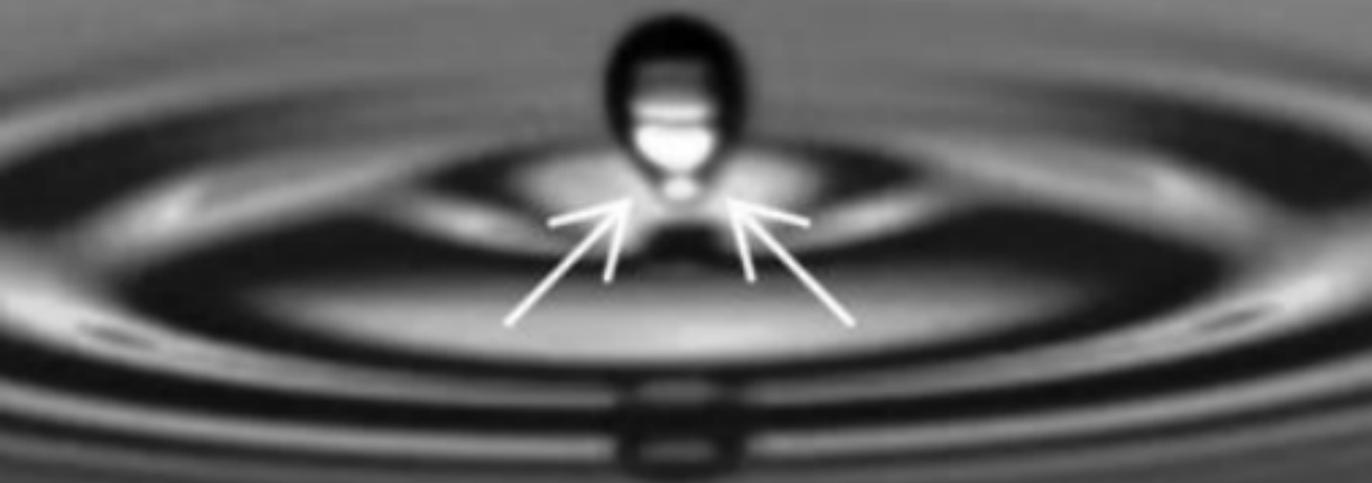

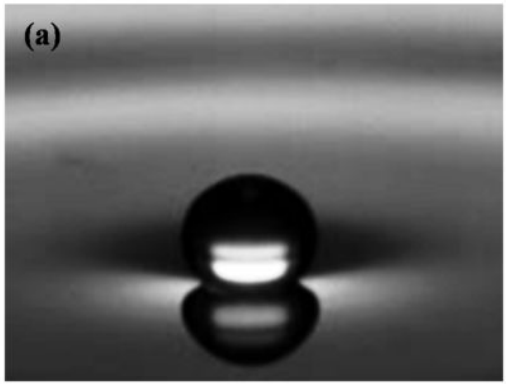
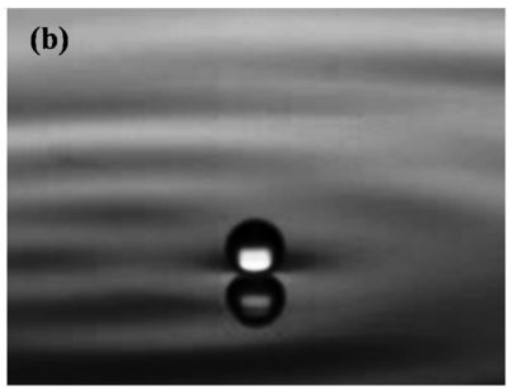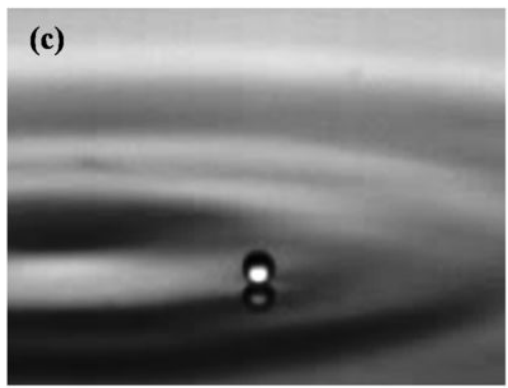
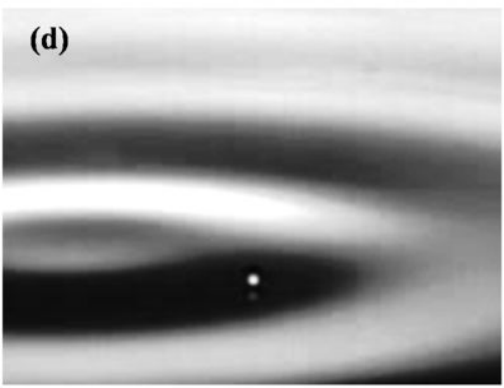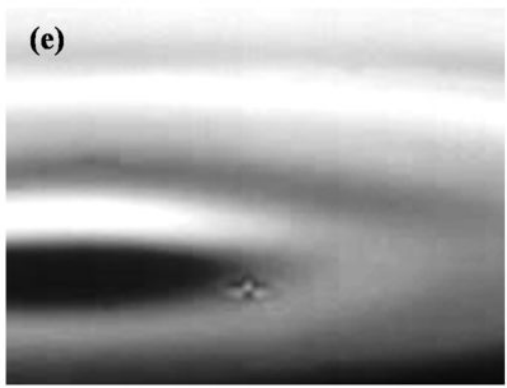

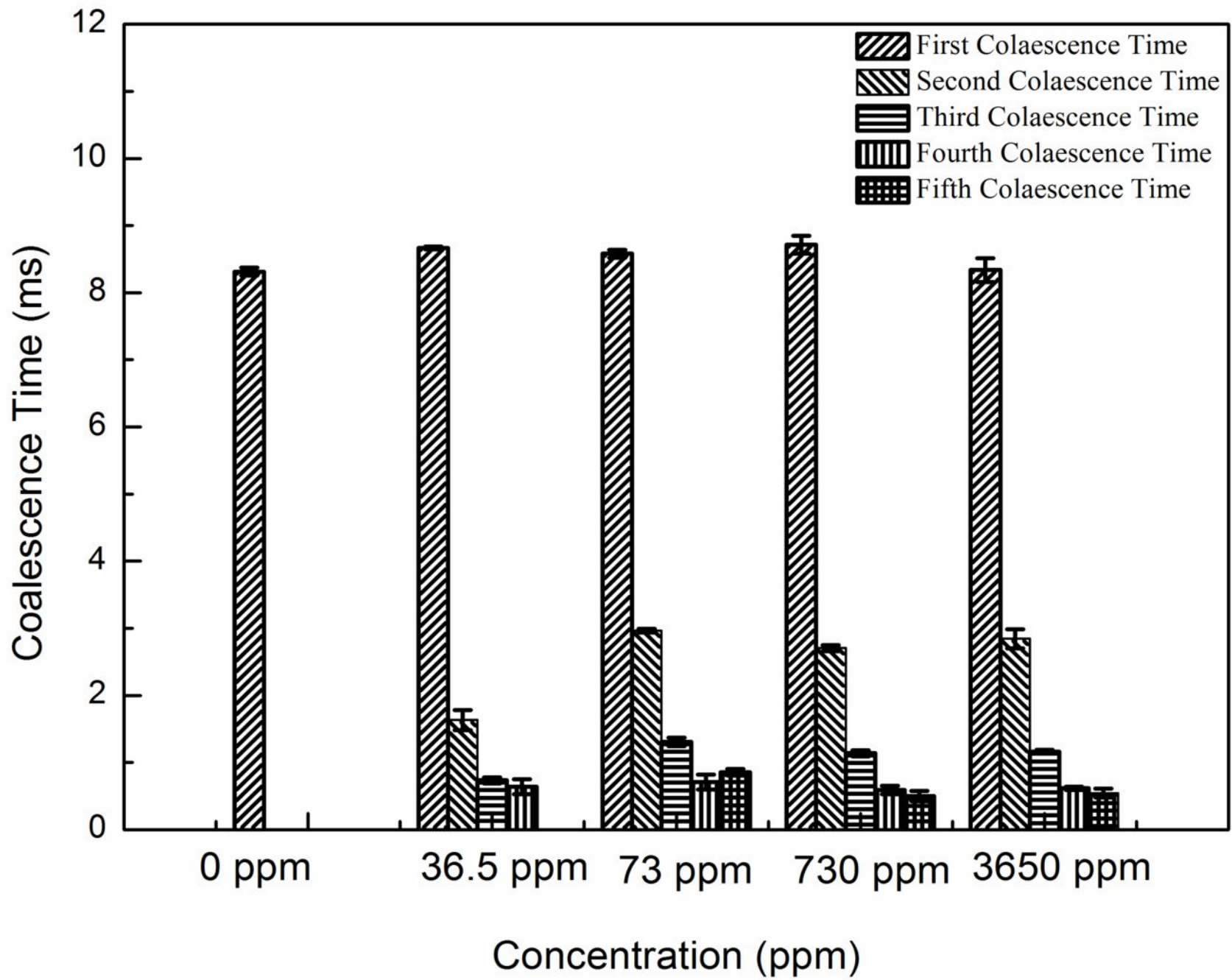